\pdfoutput=1
\documentclass{article}
\usepackage{graphicx}
\usepackage{authblk}
\usepackage{color}
\usepackage{xcolor}
\graphicspath{{figures/}}
\usepackage{amsmath,amssymb}
\usepackage{amsfonts}

\usepackage[toc, page]{appendix}

\usepackage[margin=2.0cm]{geometry}

\usepackage{float}
\usepackage{booktabs}
\usepackage{colortbl}
\definecolor{Gray}{gray}{0.55}
\definecolor{LightGray}{gray}{0.92}
\newcommand{\T}{\rule{0pt}{2.6ex}}
\newcommand{\B}{\rule[-1.2ex]{0pt}{0pt}}
\newcommand{\Tsmall}{\rule{0pt}{2.1ex}}
\newcommand{\TopT}{\rule{0pt}{3.2ex}}
\usepackage[numbers,square]{natbib}
\usepackage[colorlinks=true,linkcolor=blue,citecolor=blue,urlcolor=blue]{hyperref}

\newcounter{movie}

\newcommand{\um}{{\textmu m}}

\title{
Proliferating against the current: how multistep branching nucleation can sort microtubule polarity in the spindle
}

\author[1]{Ido Lavi}
\author[2]{Sebastian F\"urthauer}
\author[1,3,4]{Daniel J. Needleman}
\author[5,1]{Michael J. Shelley}
\affil[1]{Center for Computational Biology, Flatiron Institute, Simons Foundation, New York, NY 10010, USA}
\affil[2]{Institute of Applied Physics, TU Wien, Lehargasse 6, 1060 Vienna, Austria}
\affil[3]{Department of Molecular and Cellular Biology, Harvard University, Cambridge, MA 02138, USA}
\affil[4]{School of Engineering and Applied Sciences, Harvard University, Cambridge, MA 02138, USA}
\affil[5]{Courant Institute of Mathematical Sciences, New York University, New York, NY 10012, USA}

\begin{document}
\maketitle
\begin{abstract}
The spindle, which segregates chromosomes during cell division, is a paradigm of cytoskeletal self-organization: a highly ordered structure that persists for hours is assembled from dynamic microtubules that branch, grow, slide, and turn over within tens of seconds. Its microtubules point inward, with minus ends clustered at the poles and plus ends extending toward the chromosomes at the center of the structure. In large meiotic spindles this arrangement is puzzling, since microtubules nucleate near the chromosomes and grow at their plus ends much faster than motors slide them backward. One would therefore expect the plus ends, rather than the minus ends, to radiate outward. Here, we propose a resolution to this seeming paradox through a kinetic model that integrates the core filament-level processes. When branching complexes assemble with a sufficient delay, for instance through several sequential rate-limiting steps, they accumulate on old microtubule lattice sites near the backward-sliding minus ends of mother microtubules. Each generation is then born farther back than the last, so the self-renewing population drifts toward the minus ends even though each filament's center advances toward its plus end. For the sliding and growth velocities measured in \emph{Xenopus} egg extract spindles, a multistep mechanism requires at least four rate-limiting steps. The kinetic theory has a surprisingly amenable structure: the filament length configuration variable can be eliminated exactly, yielding a finite space-time hierarchy, and the direction of population drift follows from a general criterion comparing the mean lattice age at branching sites with the ratio of plus-end advance to sliding speed.
\end{abstract}

In the complex and crowded environment of a cell, subcellular structures must establish appropriate size, position, and internal organization to perform their biological function~\cite{levy2012mechanisms,reber2015intracellular}. Examples of cytoskeletal structures include the actin-based cortex, which maintains cell shape and mechanical integrity~\cite{salbreux2012actin}, and centrosomal asters, whose microtubule (MT) arrays organize and position the mitotic spindle~\cite{mcnally2013mechanisms}.
These filamentous structures, based on actin or MTs, are interconnected, organized by molecular motors, and grow through autocatalytic nucleation, where new filaments branch from pre-existing ones~\cite{mullins1998interaction,petry2013branching,decker2018autocatalytic}. These processes underlie their rapid and robust self-assembly.

The meiotic anastral spindle, which assembles without centrosomes in systems like the \textit{Xenopus} egg and its extract \cite{heald1996self}, exemplifies the complexity of cytoskeletal self-organization. It is built from many dynamic components: active motors that crosslink and slide filaments; branching nucleation regulated by the RanGTP pathway (a chromosome-derived signal); plus-end polymerization; and MT catastrophe.
While each process has been studied in detail, it remains
unclear how they function collectively to generate a spatially organized bipolar structure. It has been proposed that spatially regulated branching nucleation, in concert with motor-based active transport, is the key to constructing the spindle architecture~\cite{thawani2019spatiotemporal,brugues2012nucleation}, yet to what extent the specifics of the branching reaction govern the macroscopic self-organization of spindle MTs
is unknown.

Here we develop a theoretical framework that integrates these core processes into a unified kinetic description of spindle assembly.
This integrated model
points to a puzzle: MT growth is significantly faster than kinesin-driven sliding, raising the question of how slowly drifting MT minus ends, rather than rapidly polymerizing plus ends, end up radiating outwards from the center of the structure.
Filament-based simulations suggest that minus-end focusing by dynein-like motors can support the observed polarity in this regime, given a suitable binding mechanism~\cite{chew2025molecular}; yet inward polarity persists experimentally even when dynein is inhibited~\cite{burbank2007slide,dalton2022gelation}.
We propose that the resolution to this puzzle lies in the nucleation process itself. When branching requires multiple sequential binding steps of molecular factors, fully assembled branching complexes become clustered near the minus ends of MTs, which persistently drift backwards under the action of kinesin motors. 
These clustered sites then serve as platforms for nucleating new MTs which inherit the polarity of their parent. As a result, newly branched MTs are systematically offset in the minus-end direction relative to their predecessors, creating a persistent backward drift at the population level, even though individual filament centers always advance forward due to rapid plus-end growth. For $M$ sequential rate-limiting steps, this backward drift requires the ratio of sliding to growth speed to exceed $1/(M+2)$, so that at the measured ratio of about $1/5$ at least four such steps are needed. We further show that a related mechanism accomplishes the same task: cooperative self-recruitment, in which a single branching species promotes the binding of further copies of itself, produces a similar minus-end bias and can drive the requisite backward drift on its own. Both are instances of a general principle: what matters is a delay in assembly, such that branching occurs preferentially on older lattice sites that have drifted backward with the mother filament. We derive a criterion for polarity sorting that is independent of the molecular details of the delay: the mean residence time of the lattice at branching sites, in units of the MT lifetime, must exceed the ratio of the plus-end velocity to the sliding speed.

Our model is expressed in the Doi--Edwards kinetic framework \cite{doi1988theory}, describing the evolution of an MT number density whose configuration variables are MT positions, orientations, and lengths. For ease of explication, we focus here on a one-dimensional (1D) spatial system of growing MTs, pointing either right or left.
Microtubules are birthed through an autonucleation process in which new MTs branch from nucleating complexes bound to existing ones \cite{travis2022microtubules,decker2018autocatalytic}. The formation of branching complexes is modeled as a sequence of multiple binding steps, generalizing a previous study \cite{thawani2019spatiotemporal}, with a chromosome-derived signal (RanGTP) locally regulating the branching reaction \cite{petry2013branching,kraus2023augmin}. Until disassembly through catastrophe, each MT (i) grows at its plus end with a constant polymerization velocity, and (ii) is moved towards its minus end by active crosslinkers. Though restricted to 1D, this level of description contains the essential ingredients needed to address the question of how the spindle polarity structure is achieved.

A nontrivial property of our model is that filament length can be exactly marginalized, reducing the $(2+1)$-dimensional (space, length, and time) kinetic problem to a closed system in space and time alone. This reduction follows directly and exactly from the structure of the growth and nucleation processes and extends naturally to arbitrary numbers of sequential assembly steps. It also persists in higher spatial dimensions, facilitating extensions to full 3D spindle geometry and, more broadly, to other cytoskeletal architectures where branching nucleation and growth play a central role.

The paper is organized as follows. Section~\ref{sec:basic-model} introduces the integrated kinetic framework.
Section~\ref{sec:single-step} applies this framework to single-step branching nucleation and shows that it cannot reproduce the correct spindle polarity at measured velocities of MT growth and sliding. 
Section~\ref{sec:multistep} extends the model to multistep branching, demonstrating that a minimum of four rate-limiting assembly steps resolves the polarity-sorting puzzle. Section~{\ref{sec:generalization}} shows that these results, together with those of the cooperative model (Appendix~{\ref{app:cooperative}}), are special cases of a general criterion for polarity sorting by delayed assembly.

\section{A unified kinetic model}
\label{sec:basic-model}

\begin{figure}[tp]
\begin{center}
\includegraphics[scale=0.7]{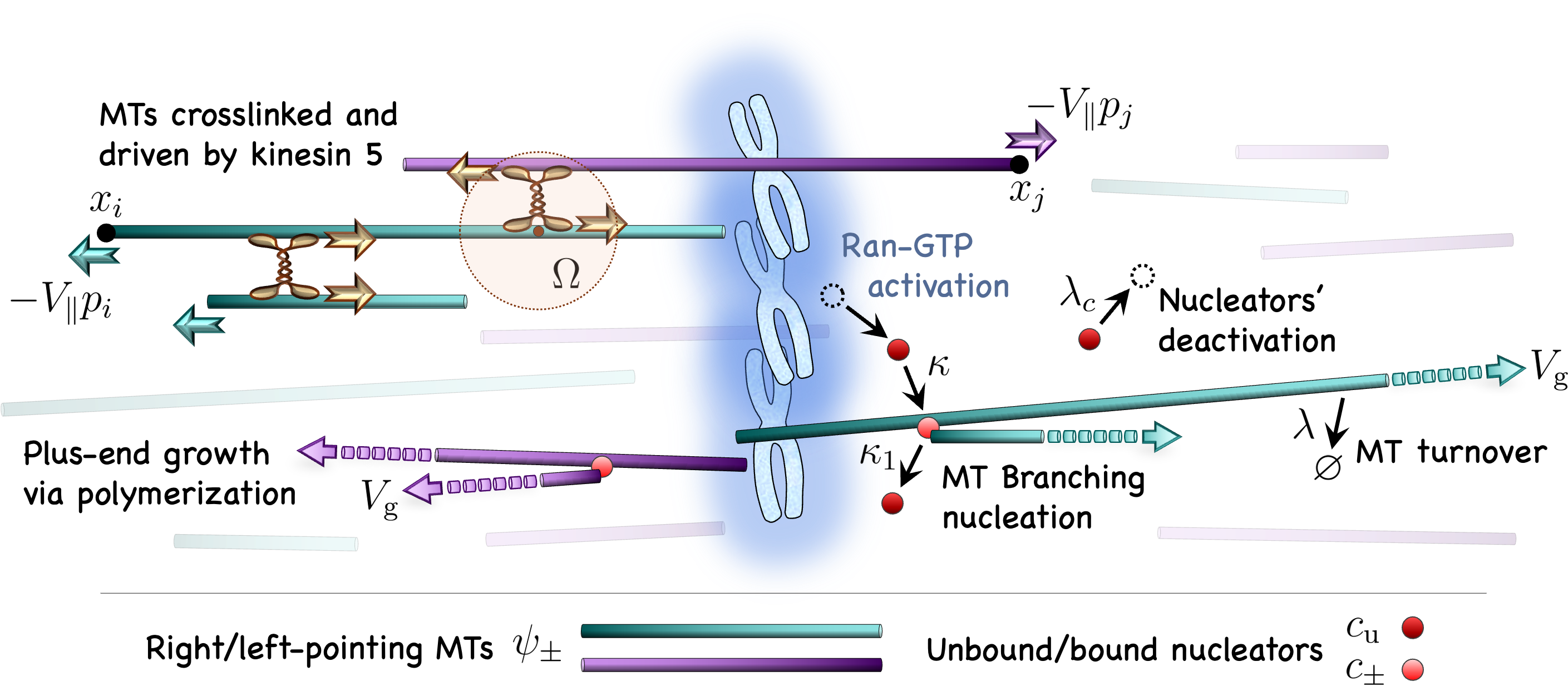}
\end{center}
\caption{{\bf Schematic of the unified model with single-step branching nucleation.}
Right-pointing (teal) and left-pointing (purple) MTs are crosslinked by active kinesin-5 motors (gold), which slide antiparallel filaments apart within the interaction zone $\Omega$.
In the highly crosslinked regime, each MT slides back at velocity $-V_\parallel$ relative to its polarity $p_i=\pm1$. Once nucleated, MTs continuously grow at velocity $V_\text{g}$ from their plus end and turn over at rate $\lambda$. Branching nucleation is controlled by a reaction-diffusion process: chromosomes at the center of the domain ($x=0$) emit a RanGTP signal $s(x)$ (blue haze) that activates unbound nucleators (red) with concentration $c_\text{u}$.
These diffuse around and bind at rate $\kappa$ to existing MTs, giving rise to the bound populations $c_\pm$ (light red). The latter trigger branching nucleation at rate $n_{\pm}$ and unbind at rate $\kappa_1$. All nucleators are deactivated at rate $\lambda_c$.
}
\label{Fig schematic basic}
\end{figure}

Experiments on \textit{Xenopus} egg extract spindles reveal that two essentially antiparallel MT populations emerge at steady state \cite{heald1996self,tirnauer2004microtubule,yang2008regional,needleman2010fast,burbank2007slide,dalton2022gelation}. In the absence of dynein motors, members of these anti-aligned populations are observed to flow in opposite directions at a roughly constant speed of
$\sim2$--$3\,$\um/min, independent of the local polarity or filament overlap \cite{burbank2007slide,dalton2022gelation}. This behavior is consistent with the theory of highly crosslinked filament ensembles \cite{furthauer2019self,furthauer2021design}, where the persistent backward motion of filaments relative to their own orientation reflects a local force balance within the dense network, generated by many multimeric, plus-end–directed motors. Within this context, it is reasonable to reduce the problem to two opposing MT orientations that branch, grow and translate in a single spatial dimension, aligned with the spindle axis, as illustrated in Fig.~\ref{Fig schematic basic}. Such a reduction assumes that nematic order is well established, as has been observed experimentally \cite{heald1996self,yang2008regional,needleman2010fast,brugues2014physical,dalton2022gelation} and is expected from steric excluded-volume interactions in dense rod suspensions \cite{doi1988theory}.
Since polarity sorting concerns the arrangement of plus and minus ends along the spindle axis, this reduction retains the dynamics that determine it. Our model assumes invariance in the transverse directions, and that the nucleation-activating signal (RanGTP) is emitted symmetrically left and right from a point source (chromosomes) at the origin of the domain; see Fig.~\ref{Fig schematic basic}.

Let $\psi_\pm(x,l,t)$ denote the number densities of right-pointing ($+$) and left-pointing ($-$) \emph{growing} filaments of length $l$, with their \emph{minus ends} located at position $x$ at time $t$. For reference, the plus ends of such filaments are located at $x\pm l$, respectively, for the ($\pm$) populations. The number densities obey the Fokker--Planck equations:

\begin{equation}
    \partial_t\psi_{\pm}(x,l,t) \mp V_\parallel\partial_x\psi_{\pm}+V_\text{g}\partial_l\psi_{\pm}=-\lambda\psi_{\pm}, \label{eq kinetic growth}
\end{equation}
where $\lambda$ is the MT catastrophe rate, $V_\text{g}$ is the growth velocity of MT plus ends, and $V_\parallel$ is the sliding speed of individual MTs, taken as a constant and directed opposite to the MT polarity. This sliding arises from crosslink-mediated interfilament interactions in a dense network driven by kinesin-5–type motors (multimeric and plus-end–directed); see Appendix~\ref{app:force-balance}.

The nucleation of new filaments, at zero length, is introduced as a flux at the boundary $l=0$, 
\begin{equation}
V_\text{g}\psi_\pm|_{l=0}=n_\pm(x,t), \label{eq flux condition}
\end{equation}
with $n_\pm(x,t)$ being the rate at which new filaments of orientation ($\pm$) are nucleated at position $x$ and time $t$. As we explicitly model branching, the nucleation rate depends nontrivially on $\psi_\pm$. Since nucleation occurs at $l = 0$, and $V_\text{g}$ is finite, turnover ensures that $\psi_\pm \to 0$ as $l \to \infty$. Moreover, for spatially localized $n_\pm$ and finite $V_\parallel$, one also has $\psi_\pm \to 0$ as $x \to \pm\infty$.
As shrinkage rates are much faster than growth, and recovery from catastrophe is rare, our kinetic model only accounts for growing MTs which, upon catastrophe, vanish from the population. 

Equation~\eqref{eq kinetic growth} is a constant-coefficient linear partial differential equation (PDE). The only explicit spatial dependence comes through $n_\pm(x,t)$ in the flux condition Eq.~\eqref{eq flux condition}. Hence, given the nucleation rates $n_\pm(x,t)$, the solution follows directly from the method of characteristics, as detailed in Appendix~\ref{app:characteristics}. 
 With initial conditions set by $\psi_{\pm,0}(x,l)$ at $t=0$, the solution reads 
 \begin{equation}
 \psi_{\pm}(x,l,t) = \begin{cases}
    \psi_{\pm,0}\left(x \pm V_\parallel t,\, l-V_\text{g} t\right) e^{-\lambda t}, & l\geq V_\text{g} t \\
\frac{1}{V_\text{g}} n_{\pm}\left(x \pm \tfrac{V_\parallel}{V_\text{g}} l,\, t-\tfrac{1}{V_\text{g}}l\right) e^{-\lambda l/V_\text{g}},\, & l < V_\text{g} t 
 \end{cases}  , \label{eq:characteristics}
\end{equation}
where the first branch accounts for MTs that already existed at $t=0$ (set by the initial condition) and the second for those born after $t=0$ via $n_\pm$. Equation~\eqref{eq:characteristics} expresses $\psi_\pm$ in terms of $n_\pm$ and its history, but does not close the problem: $n_\pm$ is itself a functional of $\psi_\pm$. In practice, $n_\pm$ is evolved as part of the full time-dependent system. At steady state, however, history drops out and $\psi_\pm$ can be reconstructed from the steady $n_\pm(x)$ alone; see Appendix~\ref{app:characteristics}.

\section{Single-step branching nucleation}
\label{sec:single-step}
We first consider a model in which branching MT nucleation results from a nucleator that triggers branching in a single step. In this model, the nucleation rate $n_\pm(x,t)$ depends on the local concentration of MT-bound nucleators, which are activated by an imposed, spatially varying signal (RanGTP, emitted from the chromosomes). 
We denote by $c_\text{u}(x,t)$ the concentration of the unbound (diffusing) nucleators, and by $c_\pm(x,t)$ the concentration of nucleators bound to polymerized tubulin along right-pointing ($+$) and left-pointing ($-$) MTs. We emphasize that $c_\pm(x,t)$ represents the concentration of bound nucleators at the laboratory coordinate $x$, irrespective of where they are bound relative to the minus or plus ends of MTs. 

Branching MTs are nucleated at a small angle relative to their parent filament~\cite{petry2013branching,kamasaki2013augmin,thawani2019spatiotemporal,gouveia2023acentrosomal}, with angle median $0^\circ$ and standard deviation $9^\circ$ measured in \textit{Xenopus} egg extracts~\cite{thawani2019spatiotemporal}, so daughter MTs predominantly inherit the polarity of their mother.
In the 1D setting with two opposing orientations, we
assume strictly parallel nucleation. The nucleation rate is given by $n_\pm(x,t)=n_bc_\pm(x,t)$, where $n_b$ is the rate at which each bound nucleator triggers a branching event.

We denote by $s(x)$ the fixed, normalized profile of RanGTP, which activates the nucleators (see Fig.~\ref{Fig schematic basic}). We take it to decay exponentially from the origin, $s(x)=e^{-|x|/\ell_s}$, as expected for the steady-state diffusion gradient of a species produced at a point source and deactivated in the cytoplasm. The signal enters through a saturating activation filter, $f_\text{act}(s)=a_c\, s/(s_0+s)$, rather than linearly. This assumes that activation draws on a finite cytoplasmic reservoir, e.g., a limited pool of inactive nucleators, or limited activating machinery. The response is linear where the signal $s$ is weak and saturates at $a_c$ where $s\gg s_0$. This nonlinearity is not essential to the polarity-sorting mechanism itself; its role is to keep the activated nucleator concentration nearly flat across a broad midzone, consistent with the broad nucleation zone inferred from measurements (Section~\ref{sec:single-step-failure}).
The transport and kinetics of the nucleator species are given by:
\begin{align}
    &\partial_t c_\text{u}(x,t)-D\partial_x^2c_\text{u}=-\kappa\left(\phi_+^{(0)}+\phi_-^{(0)}\right)c_\text{u}+\kappa_1(c_++c_-)-\lambda_c c_\text{u}+f_\text{act}\left(s(x)\right), \\
    &\partial_t c_\pm(x,t) \mp V_\parallel\partial_x c_\pm  =\kappa\,\phi_\pm^{(0)} c_\text{u}-\kappa_1 c_\pm-\lambda_c c_\pm,
\end{align}
where $D$ is the diffusion coefficient of the unbound molecules, $\kappa$ is their binding rate to polymerized tubulin along MTs, $\kappa_1$ is the unbinding rate (accounting also for MT catastrophe at rate $\lambda$), and $\lambda_c$ is the nucleator deactivation rate (whether bound or unbound). In addition, we introduce $\phi_\pm^{(0)}(x)$ as the density of potential binding sites for $c$ on the tubulin lattice present at $x$. It is computed as the number of filaments with orientation $(\pm)$ having material points at position $x$, normalized by a molecular length scale $\ell_0$ representing the effective size of a binding site:
\begin{equation}
    \phi_\pm^{(0)}(x) =\frac{1}{\ell_0}\int_0^\infty dl\int_0^lds\,\psi_\pm(x\mp s,l)=:\mathcal{L}_\pm^{(0)}\left[\psi_\pm\right]. \label{eq phi_pm}
\end{equation}
The operator $\mathcal{L}^{(0)}_\pm[\psi_\pm]$ sums contributions from all filaments of a given polarity passing through $x$, weighting each equally regardless of minus-end position or length. The inner integral (over $s$) accounts for all minus-end positions of filaments of length $l$ that cross $x$, and the outer integral (over $l$) sums over all possible lengths. The convective velocity of $\phi_\pm^{(0)}$ (and $c_\pm$ by association) is simply $\mp V_\parallel$.

We define the total concentration of nucleators as $c=c_\text{u}+c_++c_-$. By adding up the transport equations we obtain 
\begin{equation}
    \partial_t c(x,t)+V_\parallel\partial_x(c_- -c_+)-D\partial_x^2 c_\text{u}=-\lambda_c c+f_\text{act}(s). \label{eq: c no approx}
\end{equation}
If we now assume that the kinetic rates ($\kappa$, $\kappa_1$) are much faster than the rates of transport in the system, and the deactivation rate $\lambda_c$, we can take the quasi-static approximation \cite{oh2016spatial,kaye2018measuring},
\begin{equation}
    \frac{c_\pm}{c_\text{u}} \simeq \alpha \phi_\pm^{(0)},
    \quad \text{where } \alpha = \frac{\kappa}{\kappa_1}.
\end{equation}
Formally, this corresponds to
\begin{equation}
c_\text{u} = \frac{c}{1 + \alpha \left(\phi_+^{(0)} + \phi_-^{(0)}\right)} + \delta c_\text{u}, 
\quad 
c_\pm = \frac{\alpha \phi_\pm^{(0)} c}{1 + \alpha \left(\phi_+^{(0)} + \phi_-^{(0)}\right)} + \delta c_\pm,
\end{equation}
where $\delta c_\text{u},\delta c_\pm\ll c$. Then, to leading (zeroth) order in the deviations $\delta c_\text{u}$ and $\delta c_\pm$, Eq.~\eqref{eq: c no approx} reduces to a closed-form equation for $c$:
\begin{equation}
    \partial_t c(x,t) 
    + V_\parallel\partial_x\left( \frac{\alpha\, \left(\phi_-^{(0)}-\phi_+^{(0)}\right)}{1 + \alpha\left(\phi_+^{(0)} + \phi_-^{(0)}\right)} c\right) 
    - \, \partial_x^2 \left( \frac{D}{1 + \alpha\left(\phi_+^{(0)} + \phi_-^{(0)}\right)} c \right)
    = -\lambda_c c + f_\text{act}(s).
    \label{eq: c approx}
\end{equation}
Finally, as a function of $\phi_\pm^{(0)}$ and $c$, the nucleation rate is given by
\begin{equation}
n_\pm(x,t)=n_b c_{\pm}(x,t)\simeq n_b\frac{\alpha\,\phi_\pm^{(0)}(x,t)\, c(x,t)}{1+\alpha\left(\phi_+^{(0)}(x,t)+\phi_-^{(0)}(x,t)\right)}. \label{eq n single}
\end{equation}
Through its dependence on $c$, itself sourced by $f_\text{act}(s)$, this expression localizes autocatalytic nucleation to the Ran-activated midzone. In addition, the denominator saturates the nucleation rate at high binding-site densities $\phi^{(0)}_\pm$. When binding sites are abundant, the bound nucleator pool $c_\pm$ is spread across more tubulin lattice sites, reducing the per-site availability. In this single-step model, the nucleation rate per lattice site is proportional to the available nucleator concentration $c_\text{u}$. If it is uniform along a mother MT, branching occurs with equal probability per unit arclength.

\subsection{The single-step reduced dimension problem}
\label{sec:reduced-dim-single}
The model for $\psi_\pm(x,l,t)$ given by Eqs.~\eqref{eq kinetic growth} and~\eqref{eq flux condition}, with $\phi_\pm^{(0)}(x, t)$ in Eq.~\eqref{eq phi_pm} and $c(x, t)$ in Eq.~\eqref{eq: c approx} setting the nucleation rate via Eq.~\eqref{eq n single}, is posed in 2+1 dimensions: $x,l,t$. This problem conveniently admits an equivalent formulation independent of the length variable $l$. Rather than solving for the full number densities, $\psi_\pm(x,l,t)$, we derive a dynamical equation directly for $\phi_\pm^{(0)}(x, t)$, which are the fields that feed into $n_\pm$. 
Applying $\mathcal{L}_\pm^{(0)}$ of Eq.~\eqref{eq phi_pm} to Eq.~\eqref{eq kinetic growth} yields, term by term:
\begin{align}
    &\mathcal{L}_\pm^{(0)}[\partial_t\psi_\pm]=\partial_t\mathcal{L}_\pm^{(0)}[\psi_\pm]=\partial_t\phi^{(0)}_\pm, \\
    &\mathcal{L}_\pm^{(0)}[\mp V_\parallel\partial_x\psi_\pm]=\mp V_\parallel\partial_x\mathcal{L}_\pm^{(0)}[\psi_\pm]=\mp V_\parallel\partial_x\phi^{(0)}_\pm, \\
    &\mathcal{L}_\pm^{(0)}[V_\text{g}\partial_l\psi_\pm]=\frac{1}{\ell_0}\int_0^\infty dl\int_0^lds \,V_\text{g} \partial_l\psi_\pm(x\mp s,l) \\
    & \qquad\quad =\frac{V_\text{g}}{\ell_0}\int_0^\infty ds\int_{s}^\infty dl\,\partial_l\psi_\pm(x\mp s,l)  =\frac{V_\text{g}}{\ell_0} \int_0^\infty ds \left[\psi_\pm(x\mp s,l)\right]^{l=\infty}_{l=s}\nonumber \\
    & \qquad\quad =-\frac{V_\text{g}}{\ell_0}\int_0^\infty ds \,\psi_\pm(x \mp s,s). 
    \nonumber
\end{align}
The last integral sums over all MTs with their plus end located at position $x$ (filaments with minus end at $x\mp s$ and length $s$), which we identify as the density of MT plus ends:
\begin{equation}
    \eta_\pm(x,t)=\int_0^\infty \psi_\pm(x\mp l,l,t)dl=:\mathcal{G}_\pm[\psi_\pm]. \label{eq eta_pm}
\end{equation}

We then have
\begin{equation}
    \partial_t\phi^{(0)}_\pm(x,t)\mp V_\parallel\partial_x\phi^{(0)}_\pm=\frac{V_\text{g}}{\ell_0}\eta_\pm-\lambda\phi^{(0)}_\pm. \label{eq phi pm 0}
\end{equation}
As expected, the densities $\phi^{(0)}_\pm$ of polymerized tubulin lattice sites are advected as MT material points with velocities $\mp V_\parallel$ and are lost at the global catastrophe rate $\lambda$. The source term, appearing first on the right-hand side (RHS), also has a natural interpretation: each polymerizing plus end continuously lays down new lattice sites and thereby adds to $\phi_\pm^{(0)}$ at rate $V_\text{g}/\ell_0$.

To close the problem, we must also specify the dynamics of $\eta_\pm$. To that end, we apply $\mathcal{G}_\pm$ of Eq.~\eqref{eq eta_pm} to Eq.~\eqref{eq kinetic growth}, term by term:
\begin{align}
    &\mathcal{G}_\pm[\partial_t\psi_\pm]=\partial_t\mathcal{G}_\pm[\psi_\pm]=\partial_t\eta_\pm, \\
    &\mathcal{G}_\pm[\mp V_\parallel\partial_x\psi_\pm]=\mp V_\parallel\partial_x\mathcal{G}_\pm[\psi_\pm]=\mp V_\parallel\partial_x\eta_\pm, \\
    &\mathcal{G}_\pm[V_\text{g}\partial_l\psi_\pm]=V_\text{g}\int_0^\infty ds\,[\partial_l\psi_\pm](x\mp s,s)=V_\text{g}\int_0^\infty ds\,\partial_2\psi_\pm(x\mp s,s) \\
    &\qquad \quad =V_\text{g}\int_0^\infty ds\left[\frac{d}{ds}\pm \partial_1\right]\psi_\pm(x\mp s,s)\nonumber\\
    &\qquad \quad =V_\text{g}\left[\psi_\pm (x\mp s,s)\right]^{s=\infty}_{s=0}\pm V_\text{g}\mathcal{G}_\pm[\partial_x\psi_\pm] =-n_\pm(x)\pm V_\text{g}\partial_x \eta_\pm,\nonumber
\end{align}
where we used the fact that 
$\frac{d}{ds}\psi_\pm(x\mp s,s)=\mp\partial_1\psi_\pm(x\mp s,s)+\partial_2\psi_\pm (x\mp s,s),$
and substituted the boundary condition at $l=0$, given by Eq.~\eqref{eq flux condition}.

We then have
\begin{equation}
    \partial_t\eta_\pm(x,t)\pm(V_\text{g}- V_\parallel)\partial_x\eta_\pm =n_\pm-\lambda\eta_\pm. \label{eq eta pm} 
\end{equation}
This equation correctly reflects that, rather than moving as ordinary material points, the plus ends of right/left-pointing filaments advance with an effective velocity $\pm(V_\text{g}-V_\parallel)$. The loss term due to catastrophe is also expected, and the source term corresponds to the filament nucleation rate $n_\pm(x,t)$. By definition, newly nucleated filaments start with zero length, so their minus and plus ends initially coincide at the same nucleation position $x$.

Taken together, Eqs.~\eqref{eq phi pm 0}, \eqref{eq eta pm}, \eqref{eq: c approx}, and \eqref{eq n single} form a closed, \emph{length-independent} dynamical system for the five fields: $\phi^{(0)}_\pm(x,t)$, $\eta_\pm(x,t)$, and $c(x,t)$.

\subsection{The single-step model fails to predict correct spindle polarity using measured velocities}
\label{sec:single-step-failure}
Unlike the parameters of branching nucleation, which are largely unknown, three quantities that characterize MT dynamics in our model have been measured: the catastrophe rate $\lambda$, the sliding speed $V_\parallel$, and the polymerization speed $V_\text{g}$. In \textit{Xenopus} egg extract spindles, MT lifetimes are typically 15--20\,s, sliding speeds are 2--3\,\um/min, and plus ends advance at 10--30\,\um/min (Appendix~\ref{app:measured-parameters}, Table~\ref{table measured}). A typical MT thus reaches a length $\ell=V_\text{g}/\lambda\approx4$\,\um, whereas its minus end slides only about $V_\parallel/\lambda\approx1$\,\um{} over its lifetime, a small fraction of the 30--50\,\um{} spindle; the nucleation zone around the chromosomes must therefore be broad \cite{burbank2007slide}. The ratio $V_\parallel/V_\text{g}$, independent of $\lambda$, is the key determinant of polarity sorting. Taking $V_\parallel\approx2.5$ and $V_\text{g}\approx12.5$\,\um/min, we adopt $V_\parallel/V_\text{g}\approx1/5$ as our reference value, noting that reported values span a wide range.

\begin{figure}[tp]
\begin{center}
\includegraphics[scale=1]{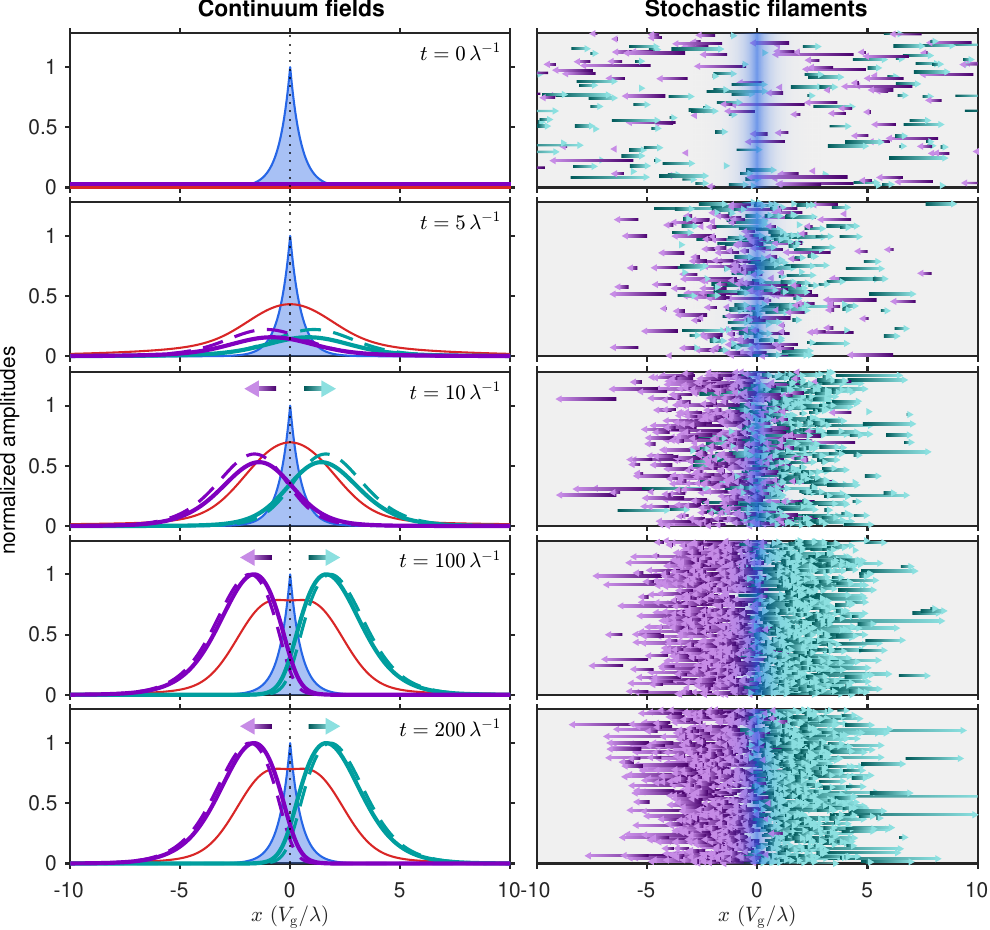}
\end{center}
\caption{{\bf Relaxation to steady state in the single-step model ($M=0$).}
Left column: time evolution of the continuum fields: Ran source $s(x)$ (blue, fixed), activated nucleator $c(x,t)$ (red), tubulin densities $\phi^{(0)}_\pm(x,t)$ (solid teal/purple), and plus-end densities $\eta_\pm(x,t)$ (dashed teal/purple), normalized by their steady-state peak values. Snapshots correspond to times $t = \{0, 5, 10, 100, 200\}\lambda^{-1}$.
Arrows at the top of the graphs indicate the polarity of the respective MT population; their arrowheads (plus ends) align with the peak of the plus-end densities $\eta_\pm$, and their lengths equal $\ell = V_\text{g}/\lambda$ for reference.
Right column: corresponding stochastic filament snapshots, with right-pointing MTs in teal and left-pointing MTs in purple.
Parameters: $M=0$, $V_\parallel/V_\text{g}=0.2$, with all other parameters as in Table~\ref{table params}. The system is initialized with a uniform seed of left/right MT populations and no active nucleators.
}
\label{fig:dynamics}
\end{figure}

\begin{figure}[tp]
\begin{center}
\includegraphics[scale=1]{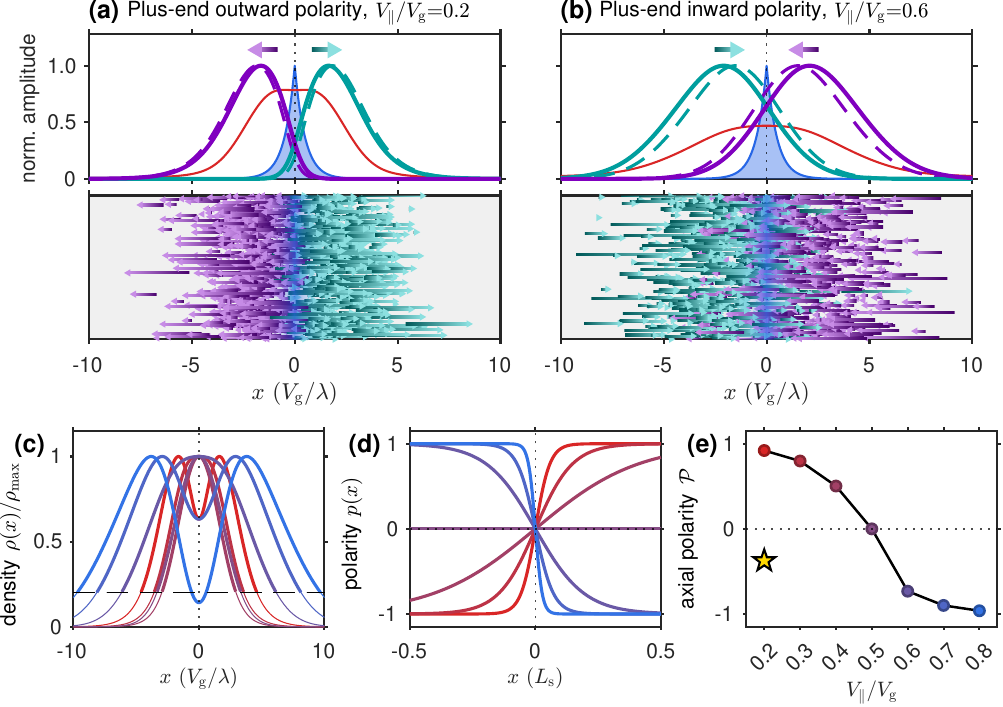}
\end{center}
\caption{{\bf The single-step model yields inverted polarity at the measured velocity ratio.}
All panels show steady-state solutions (snapshots at $t=200\lambda^{-1}$) with $M=0$ sequential stages.
\textbf{(a)} The measured velocity ratio $V_\parallel/V_\text{g}=0.2$ yields a condensate with plus-end outward (wrong) polarity.
\textbf{(b)} A higher velocity ratio of $V_\parallel/V_\text{g}=0.6$ yields plus-end inward (correct) polarity.
Top row: Ran source $s(x)$ (blue, filled) and steady-state profiles of the Ran-activated species $c(x)$ (red), MT lattice densities $\phi^{(0)}_+(x)$ (teal solid) and $\phi^{(0)}_-(x)$ (purple solid), and plus-end densities $\eta_\pm(x)$ (dashed).
Arrows at the top of the graphs indicate the polarity of the respective MT population; their arrowheads (plus ends) align with the peak of the plus-end densities $\eta_\pm$, and their lengths equal $\ell = V_\text{g}/\lambda$ for reference.
Bottom row: individual filament realizations (sampled from the steady-state $\psi_\pm(x,l)$; see Appendix~\ref{app:characteristics}), colored by the MT polarity (teal: right-pointing; purple: left-pointing), with the Ran source density overlaid as a semi-transparent blue haze.
\textbf{(c)} Normalized MT density $\rho(x)/\rho_\text{max}$, with $\rho = \phi^{(0)}_+ + \phi^{(0)}_-$, for $V_\parallel/V_\text{g} \in [0.2, 0.8]$ (red to blue); thick portion marks the spindle interior ($\rho > 0.2\,\rho_\text{max}$).
\textbf{(d)} Local polarity $p(x) = (\phi^{(0)}_+-\phi^{(0)}_-)/\rho$ within the spindle interior, $x$ normalized by spindle length $L_\text{s}$.
\textbf{(e)} Axial polarity $\mathcal{P}$ vs.\ $V_\parallel/V_\text{g}$; $\mathcal{P}>0$ indicates wrong polarity (plus ends outward), $\mathcal{P}<0$ correct polarity (plus ends inward). The star marks the experiment: at the measured velocity ratio $V_\parallel/V_\text{g}\approx 0.2$, the observed axial polarity is $\mathcal{P}\approx-3/8$, a rough estimate based on Figure~1(g) of Dalton et al.~\cite{dalton2022gelation}.}
\label{fig:M0-fields}
\end{figure}

We numerically integrate the dimensionless form of the length-independent system using a pseudo-spectral scheme described in Appendix~\ref{app:numerical-scheme} (with $M=0$), with parameters as listed in Appendix~\ref{app:parameters} (Table~\ref{table params}). Starting from a small, uniform initial population of right- and left-pointing MTs, the system is evolved to a steady state. Figure~\ref{fig:dynamics} shows the transient dynamics at $V_\parallel/V_\text{g} = 1/5$ for the single-step model. For visualization, the evolving continuum fields are accompanied by stochastic filament snapshots whose nucleation events are sampled from the continuum rate $n_\pm(x,t)$ (see Appendix~\ref{app:gillespie} for details). The full time course showing $n_\pm(x,t)$ and the stochastic filaments is found in Movie~\ref{mov:M0-eps02}. In the left panels of Fig.~\ref{fig:dynamics}, the solid profiles $\phi^{(0)}_\pm(x,t)$ give the densities of lattice sites on right- and left-pointing MTs; they are the model counterpart of the left- and right-moving speckle densities measured in \emph{Xenopus} egg extract spindles (see, e.g., Figure~1(e) of Dalton et al.~\cite{dalton2022gelation}). However, unlike the experimental data, the two simulated populations settle on the opposite sides of the origin, reflecting an inverted, plus-end outward polarity. The long-time steady state is shown in Fig.~\ref{fig:M0-fields}a. 

At the measured ratio of $V_\parallel/V_\text{g}$, the single-step model consistently yields such inverted steady states.
This stands in clear contrast to the observed polarity structure of the spindle, where right-pointing (left-pointing) MTs accumulate predominantly on the left (right) side of the origin \cite{dalton2022gelation}.
This failure is readily understood: in the single-step model, branching is uniform along the mother MT, so daughter filaments are seeded, on average, at the mother filament's center of mass. In the lab frame, this center of mass advances in the plus-end direction with velocity $V_\text{g}/2-V_\parallel$. Because backward sliding is
slow compared to the speed of polymerization ($V_\parallel<V_\text{g}/2$), each generation of daughters is displaced forward relative to the mother's origin. This persistent forward drift
drives the system to a configuration with outward-pointing polarity. 

The single-step model does admit inward, spindle-like polarity, but only when $V_\parallel$ is chosen substantially larger (or $V_\text{g}$ substantially smaller) than its measured value (Fig.~\ref{fig:M0-fields}b; Movie~\ref{mov:M0-eps06}). In the regime $V_\parallel>V_\text{g}/2$, backward sliding overcomes the growth-driven advance of the filament midpoint, so the mean branching position drifts backward.
Notably, the difficulty in reproducing spindle-like bipolarity using the \emph{measured speeds} is not unique to our minimal continuum model: similar behavior has been observed in discrete particle-based simulations that also employ a single-step branching nucleation process \cite{dalton2022gelation}.

To quantify the polar organization across a steady-state structure, we introduce a structural order parameter, $\mathcal{P}$, which we refer to as the axial polarity,
\begin{equation}
    \mathcal{P} \;=\; \frac{1}{L_\text{s}}\int_{-L_\text{s}/2}^{L_\text{s}/2} p(x)\,\mathrm{sgn}(x)\,dx,
    \label{eq:bipolarity}
\end{equation}
where $p(x) = (\phi^{(0)}_+(x)-\phi^{(0)}_-(x))/\rho(x)$ is the local polarity, with $\rho(x)=\phi^{(0)}_+(x)+\phi^{(0)}_-(x)$ the tubulin density, and $L_\text{s}$ is the spindle length, defined as the spatial extent of $\{x:\rho(x)>0.2\,\rho_\text{max}\}$. This integral measures the local polarity against the sharpest conceivable profile: $\mathrm{sgn}(x)$. Since $|p(x)|\leq1$, we have $\mathcal{P}\in[-1,1]$. For reference: $\mathcal{P}=-1$ corresponds to perfect plus-end inward polarity: $p(x)=-\mathrm{sgn}(x)$ (right-pointing MTs fully to the left, left-pointing MTs fully to the right, with no overlap); $\mathcal{P}=+1$ to perfect plus-end outward polarity: $p(x)=\mathrm{sgn}(x)$ (right-pointing fully to the right, left-pointing fully to the left); and $\mathcal{P}=0$ to balanced overlapping populations. The steady-state density and polarity profiles for varying velocity ratio $V_\parallel/V_\text{g}\in[0.2,0.8]$ are shown in Figs.~\ref{fig:M0-fields}c,d and the corresponding axial polarity is shown in Fig.~\ref{fig:M0-fields}e. In the biologically relevant parameter range, the model predicts a positive $\mathcal{P}$, contrary to experimental observation. Consistent with our argument above, inversion of the axial polarity requires $V_\parallel/V_\text{g}>0.5$, more than twice the measured ratio. 
This failure motivates revisiting the branching nucleation model, as we discuss in the next section.

\section{Multistep branching nucleation}
\label{sec:multistep}

In contrast to a single-step model (Section~\ref{sec:single-step} and \cite{dalton2022gelation,gouveia2023acentrosomal}), experiments in \textit{Xenopus} egg extracts have suggested that MT branching nucleation proceeds through a coordinated, stepwise assembly of the branching complex~\cite{travis2022microtubules}, involving at least three steps: 
the sequential recruitment of TPX2, the augmin complex, and the $\gamma$-tubulin ring complex ($\gamma$-TuRC) \cite{petry2013branching,thawani2019spatiotemporal,song2018mechanism}. In this picture, TPX2 binds the mother MT before augmin \cite{thawani2019spatiotemporal}, which in turn recruits $\gamma$-TuRC to the lattice \cite{song2018mechanism}. RanGTP stimulates these processes \cite{petry2013branching}, possibly acting on both TPX2 \cite{gruss2001ran} and augmin \cite{kraus2023augmin,ustinova2023microtubule}. Together they give rise to a mature branching complex bound to an MT lattice site, with $\gamma$-TuRC (which XMAP215 can assist \cite{thawani2020transition}) serving as the template for the nascent MT branch \cite{alfaro2020biochemical}. In addition to the recruitment of these key components, the nucleation pathway may involve additional molecular intermediates. Single-molecule experiments indicate that the rate-limiting step of nucleation from $\gamma$-TuRC is the cooperative assembly of several $\alpha\beta$-tubulin dimers on the ring \cite{thawani2020transition,consolati2020microtubule,wieczorek2021biochemical} (approximately four for \textit{Xenopus} $\gamma$-TuRC \cite{thawani2020transition}, and possibly more for human $\gamma$-TuRC \cite{consolati2020microtubule}), accompanied by a partial or complete closure of $\gamma$-TuRC from its open, isolated conformation \cite{vermeulen2024asymmetry,brito2024transition,aher2024structure}. These additional binding events therefore precede the onset of sustained $\alpha\beta$-tubulin polymerization at the plus end of an established MT. Crucially for our model, it is not yet known how many of the total sequential stages required to nucleate a daughter MT are rate-limiting, in the sense that they are slow relative to MT turnover. Our model treats the effective number of such rate-limiting stages as a free parameter $M$, as detailed below.

\begin{figure}[tp]
\begin{center}
\includegraphics[scale=0.7]{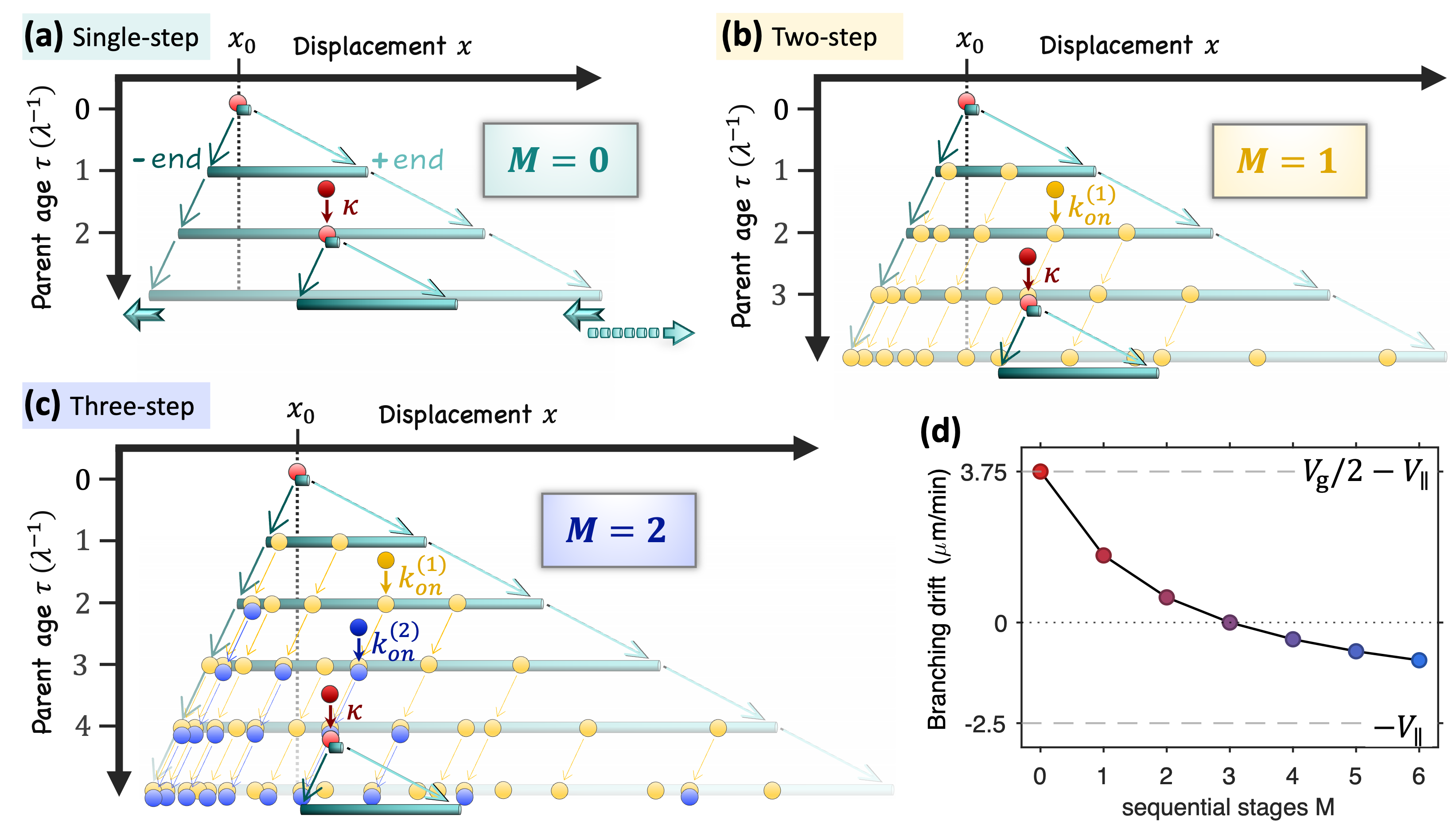}
\end{center}
\caption{{\bf Schematic of the multistep branching model and the intergenerational drift.}
\textbf{(a)}--\textbf{(c)} Illustrative space-time trajectory (kymograph) of a single right-pointing mother MT (teal), from its birth at position $x_0$ and age $\tau = 0$ through the typical branching event occurring along it, for increasing $M$. Material points along the MT, minus end included, slide backward at velocity $-V_\parallel$. Through growth, the plus end advances forward at velocity $V_\text{g}-V_\parallel$. Opacity decreases with age to indicate the survival factor $e^{-\lambda\tau}$ (rendered as a linear gradient for visibility). Each panel shows the nucleation of a daughter filament at the mean branching position and mean branching time.
\textbf{(a)} Single-step model ($M = 0$): the Ran-activated nucleator $c$ (red) binds directly to any lattice site of the mother, immediately triggering nucleation. The mean branching position is the mother's center (a distance $l/2$ from the minus end) and the mean branching time is $\tau_n = 2/\lambda$.
\textbf{(b)} Two-step model ($M = 1$ sequential stage): a first intermediate species (yellow) must bind before $c$ (red) can associate with it and trigger nucleation. Assuming bound intermediates neither unbind nor saturate, each lattice site accumulates them from the moment of its incorporation into the MT \emph{as it drifts backward}. The density is therefore highest near the minus end, where sites are oldest. The mean branching position shifts to $l/3$ from the minus end and the mean branching time shifts to $\tau_n = 3/\lambda$.
\textbf{(c)} Three-step model ($M = 2$ sequential stages): $c$ (red) associates with a second intermediate (blue) that must bind sequentially to the first (yellow) before nucleation is triggered. This further enhances the minus-end bias and lengthens the mean branching time. In general, for $M$ sequential stages, the mean position from the minus end is $l/(M+2)$ and the mean time is $\tau_n = (M+2)/\lambda$ (Section~\ref{sec:mean-branching-time}). \textbf{(d)} Each additional assembly stage reduces the forward drift of expected branching events. At the experimentally measured velocity ratio ($V_\parallel/V_\text{g}=1/5$), the drift reverses sign and becomes minus-end–directed for sufficiently large $M$ (specifically $M\geq 4$). The intergenerational drift is bounded from above by $-V_\parallel+V_\text{g}/2$ (the velocity of the mother's center) and from below by $-V_\parallel$ (the minus-end velocity).
}
\label{Fig multistep schematic}
\end{figure}

We generalize our branching nucleation model by assuming that $M\geq1$ \emph{sequential, rate-limiting} binding steps are required to form a mature branching complex on a parent MT. In this picture, assembly proceeds through a series of intermediate binding stages $m=1,2,\dots,M$ (see Fig.~\ref{Fig multistep schematic}a--c). 
Here, $M$ counts the slow assembly stages; the final, rapidly equilibrating binding of the Ran-activated species is not included in this count. The model with $M$ rate-limiting stages thus comprises $M+1$ steps in total, so that the single-step model corresponds to $M=0$.
Our model can be viewed as a continuum generalization of the discrete `sequential model' of Thawani et al.~\cite{thawani2019spatiotemporal}, which effectively maps to $M=1$ in our formalism (Fig.~\ref{Fig multistep schematic}b). 

We denote by $\phi^{(m)}_\pm(x,t)$ the density of complexes that have reached stage $m$ on right/left-pointing MTs. Our model assumes that the Ran-activated species, of concentration $c(x,t)$, rapidly binds to and unbinds from mature complexes (of densities $\phi_\pm^{(M)}$), rather than binding directly to lattice sites of polymerized tubulin (of densities $\phi_\pm^{(0)}$). Under the quasi-static approximation, the densities of $c$ unbound and bound to mature ($M$-staged) complexes on right- and left-pointing MTs reduce to:
\begin{equation}
    c_\text{u}\simeq \frac{c}{1 + \alpha \left(\phi_+^{(M)} + \phi_-^{(M)}\right)},\quad c_\pm \simeq \frac{\alpha \phi_\pm^{(M)} c}{1 + \alpha \left(\phi_+^{(M)} + \phi_-^{(M)}\right)}.
\end{equation}
Accordingly, the transport dynamics of $c(x,t)$ and the nucleation rates $n_\pm(x,t)$ are given by:
\begin{align}
    &\partial_t c(x,t) 
    + V_\parallel\partial_x\left( \frac{\alpha\, \left(\phi_-^{(M)}-\phi_+^{(M)}\right)\, c}{1 + \alpha\left(\phi_+^{(M)} + \phi_-^{(M)}\right)} \right) 
    - D\, \partial_x^2 \left( \frac{c}{1 + \alpha\left(\phi_+^{(M)} + \phi_-^{(M)}\right)} \right)
    = -\lambda_c c + f_\text{act}(s),
    \label{eq: c approx M} \\
    &n_\pm(x,t)=n_b c_\pm\simeq n_b \frac{\alpha\phi_\pm^{(M)}c}{1+\alpha\left(\phi_+^{(M)}+\phi_-^{(M)}\right)}, \label{eq n m step}
\end{align}
where we assumed that the binding of $c$ constitutes the last step in the branching reaction. Note that Eqs.~\eqref{eq: c approx M} and \eqref{eq n m step} take the same form as Eqs.~\eqref{eq: c approx} and \eqref{eq n single} of the single-step model, but with $\phi^{(M)}_\pm$ replacing $\phi_\pm^{(0)}$. 

\subsection{Distribution of branching complexes}
\label{sec:distribution}
Given the current number densities, $\psi_\pm(x, l, t)$, we postulate (and later show) that $\phi_\pm^{(m)}(x, t)$ are given by
\begin{align}
    \phi^{(m)}_+(x,t)&= \frac{A_m}{\ell_0}\int_0^\infty dl \int_0^l ds\left(\frac{l-s}{V_\text{g}}\right)^m\psi_+(x-s,l,t)=:\mathcal{L}_+^{(m)}[\psi_+]\mbox{ and}  \label{eq phi m plus} \\
    \phi^{(m)}_-(x,t)&= \frac{A_m}{\ell_0}\int_0^\infty dl \int_0^l ds\left(\frac{l-s}{V_\text{g}}\right)^m\psi_-(x+s,l,t)=:\mathcal{L}_-^{(m)}[\psi_-],  \label{eq phi m minus}
\end{align}
where
\begin{equation}
A_m=\frac{1}{m!}\prod_{m'=1}^{m} k_\text{on}^{(m')}R^{(m')},
\label{eq:Am}
\end{equation}
where $R^{(m')}$ is the uniform reservoir concentration of species $m'$, and $k_\text{on}^{(m')}$ is its binding rate to complexes assembled at the previous stage ($m'-1$). 

Hereafter, we restrict attention to the right-pointing complex density (Eq.~\eqref{eq phi m plus}, Fig.~\ref{Fig multistep schematic}), noting that the same arguments apply symmetrically to the left-pointing density (Eq.~\eqref{eq phi m minus}) with the reversed sign of polarity. In the integrand of Eq.~\eqref{eq phi m plus}, the arclength coordinate $s$ represents the distance of the point $x$ from the \emph{minus ends} of right-pointing filaments of length $l$, so $l-s$ represents the distance from $x$ to the plus ends. The inner integral (over $s\in[0,l]$) sums contributions from all such filaments intersecting the coordinate $x$, and the outer integral (over $l\in[0,\infty)$) sums over all possible MT lengths. The double integral therefore sums over all right-pointing MTs intersecting the coordinate $x$, each weighted by the factor $[(l-s)/V_\text{g}]^m$. The quantity $(l-s)/V_\text{g}$ represents the residence time of the polymerized tubulin segment present at $x$. This residence time equals $l/V_\text{g}$ at the minus end ($s=0$) and zero at the plus end ($s=l$), where tubulin dimers are freshly incorporated. Raising it to the power $m$ comes from the assumption that $m$ sequential, rate-limiting steps are needed to form the $m$-staged complex. 

In Appendix~\ref{app:rate-limiting}, we show that the specific form of Eqs.~\eqref{eq phi m plus},~\eqref{eq phi m minus} is valid under the assumptions 
\begin{equation}
k_\text{off}^{(m)}\ll \lambda \quad \& \quad k_\text{on}^{(m)}R^{(m)}\ll\lambda \quad \text{for }m\geq1,
\end{equation}
where $k_\text{off}^{(m)}$ is the unbinding rate of species $m$.
In other words, it is assumed that the conversion rates in the sequential reaction chain are all small compared with the global catastrophe rate, hence the term ``rate-limiting.'' In this regime, branching complexes accumulate along the MT over its lifetime effectively without unbinding or reaching equilibrium at any stage of assembly, so that their concentration increases with the residence time of the binding sites to which they attach. 

Equations~\eqref{eq phi m plus},~\eqref{eq phi m minus} imply that the more assembly steps $M$ required to form mature complexes, the more strongly their distribution is biased toward the minus end of the parent MT (Fig.~\ref{Fig multistep schematic}a--c). This bias is the key to polarity sorting: since daughter filaments nucleate at the location of the branching complex, it potentially translates into a backward intergenerational displacement that can overcome the forward push of plus-end polymerization (Fig.~\ref{Fig multistep schematic}d). A further consequence is that longer, older filaments host disproportionately more mature complexes, with the total count scaling as $l^{M+1}$.

\subsection{Drift of successive branching events}
\label{sec:drift}
The minus-end bias of branching complexes has a direct consequence for the intergenerational drift of the MT population. To quantify this, we track the expected position of branching events occurring along a right-pointing parent filament nucleated at position $x_0$ at time $t_0$. At age $\tau$, assuming the filament persists, its minus end is at $X(\tau) = x_0 - V_\parallel\tau$ and its length is $L(\tau) = V_\text{g}\tau$ (Fig.~\ref{Fig multistep schematic}a). Since the density of $M$-staged complexes is proportional to $(L-s)^M$ (Eq.~\eqref{eq phi m plus}), the center of mass of future branching events along the filament, assuming that the available nucleator concentration $c_\text{u}(x)$ is uniform along the parent MT, is
\begin{equation}
X^{(M)} -X= \frac{\int_0^L s(L - s)^M\, ds}{\int_0^L (L - s)^M\, ds} = \frac{L}{M + 2},
\end{equation}
which gives
\begin{equation}
X^{(M)}(\tau)= x_0 + \left(-V_\parallel + \frac{V_\text{g}}{M + 2}\right)\tau.
\end{equation}
The velocity of $X^{(M)}$ is the average drift of nucleation events across successive generations (see Fig.~\ref{Fig multistep schematic}d). This drift is directed toward the parent's minus end when:
\begin{equation}
V_\parallel > \frac{V_\text{g}}{M + 2}.
\end{equation}
For the single-step model ($M=0$, Section~\ref{sec:single-step}, Fig.~\ref{Fig multistep schematic}a), the condition reduces to $V_\parallel > V_\text{g}/2$, which, as stated already, is incompatible with the experimentally measured velocities. With the measured ratio of $V_\parallel/V_\text{g} \approx 1/5$, the condition is satisfied if $M \geq 4$: \emph{a minimum of four rate-limiting assembly steps is required for the multistep branching mechanism to support proper spindle polarity sorting}. This threshold is sensitive to the velocity ratio: a higher $V_\parallel/V_\text{g}$ would lower the required $M$.
Interestingly, for $V_\text{g}/(M+2) < V_\parallel < V_\text{g}/2$, which is likely the experimental regime (with the true $M$ to be determined), the centers of individual MTs persistently advance in the direction of their plus ends (since they grow more than twice as fast as they slide), but the self-renewing autocatalytic population as a whole nonetheless drifts in the minus-end direction (since daughters preferentially branch off old lattice sites that have already slid behind their mother's birthplace). The extension of this analysis to spatially varying $c_\text{u}(x)$ is given in Appendix~\ref{app:gradient-drift}.

\subsection{The multistep reduced-dimension problem}
\label{sec:reduced-dim-multistep}
As in the single-step model, we seek and find an $l$-independent formulation: rather than solving for the full number densities $\psi_\pm(x,l,t)$, we derive a closed dynamical system for the fields $\phi_\pm^{(m)}(x,t)$ across all stages $m\leq M$. This exact marginalization of filament length as an independent variable not only simplifies analysis and numerical integration, but also provides a self-consistent justification for the Ansatz given in Eqs.~\eqref{eq phi m plus} and~\eqref{eq phi m minus}. For the right-pointing population, we begin by applying $\mathcal{L}_+^{(m)}$ of Eq.~\eqref{eq phi m plus} to Eq.~\eqref{eq kinetic growth} term by term. To wit:
\begin{align}
    &\mathcal{L}_+^{(m)}[\partial_t\psi_+]=\partial_t\mathcal{L}_+^{(m)}[\psi_+]=\partial_t\phi^{(m)}_+, \\
    &\mathcal{L}_+^{(m)}[-V_\parallel\partial_x\psi_+]=-V_\parallel\partial_x\mathcal{L}_+^{(m)}[\psi_+]=-V_\parallel\partial_x \phi^{(m)}_+, \\
    &\mathcal{L}_+^{(m)}[V_\text{g}\partial_l\psi_+]= \frac{A_m}{\ell_0} \int_0^\infty dl\int_0^lds\left(\frac{l-s}{V_\text{g}}\right)^m V_\text{g}\partial_l\psi_+(x-s,l) \\
    & \qquad\quad =\frac{A_m}{\ell_0} \int_0^\infty ds\int_{s}^\infty dl \frac{(l-s)^m}{V_\text{g}^{m-1}}\partial_l\psi_+(x-s,l) \nonumber \\
    & \qquad\quad =\frac{A_m}{\ell_0} \int_0^\infty ds\Bigg( \left[\frac{(l-s)^{m}}{V_\text{g}^{m-1}}\psi_+(x-s,l)\right]^{l=\infty}_{l=s}  
    -m\int_{s}^\infty dl \left(\frac{l-s}{V_\text{g}}\right)^{m-1} \psi_+(x-s,l)\Bigg) \nonumber \\
    & \qquad\quad  
    =-k_\text{on}^{(m)}R^{(m)}\frac{A_{m-1}}{\ell_0} \int_0^\infty dl\int_{0}^l ds \left(\frac{l-s}{V_\text{g}}\right)^{m-1}\psi_+(x-s,l) \nonumber \\
    & \qquad \quad = -k_\text{on}^{(m)}R^{(m)}\mathcal{L}_+^{(m-1)}[\psi_+] = -k_\text{on}^{(m)}R^{(m)}\phi^{(m-1)}_+. \nonumber
\end{align}
This yields, for $m\geq1$,
\begin{equation}
    \partial_t\phi^{(m)}_+(x,t)-V_\parallel\partial_x\phi^{(m)}_+=k_\text{on}^{(m)}R^{(m)}\phi^{(m-1)}_+-\lambda\phi^{(m)}_+. \label{eq phi plus m}
\end{equation}
This equation states that species $m$, with reservoir concentration $R^{(m)}$, binds at rate $k_\text{on}^{(m)}$ to the preceding species $(m-1)$ on the branching complex. It is assumed that complexes neither slide along nor unbind from the MT; rather, they are advected with the filament material points at velocity $-V_\parallel$ and are only lost due to MT catastrophe.

From Eq.~\eqref{eq phi plus m}, the field $\phi_+^{(1)}$, representing the density of the first bound species in the sequence (possibly TPX2), is sourced by $\phi_+^{(0)}$, the bare density of polymerized tubulin lattice sites on right-pointing MTs. This field is in turn sourced by the plus-end density $\eta_+$, which is itself sourced by the nucleation rate $n_+$. The evolution equations for $\phi_+^{(0)}$ and $\eta_+$ are unchanged from the single-step model (Eqs.~\eqref{eq phi pm 0} and~\eqref{eq eta pm}), with the sole difference that the nucleation rate is now set by $\phi_+^{(M)}$ via Eq.~\eqref{eq n m step} rather than by $\phi_+^{(0)}$ via Eq.~\eqref{eq n single}. These reduced dynamics are hierarchical: the assembly stages form a feed-forward cascade, $\eta_\pm \to \phi^{(0)}_\pm \to \phi^{(1)}_\pm \to \cdots \to \phi^{(M)}_\pm$, in which each field is sourced only by the one preceding it. The cascade closes into an autocatalytic loop, since $\phi^{(M)}_\pm$ set the nucleation rates $n_\pm$ that regenerate $\eta_\pm$.

For the left-pointing population, we follow the same line of derivation to obtain:
\begin{equation}
\partial_t\phi^{(m)}_-(x,t)+V_\parallel\partial_x\phi^{(m)}_-=k_\text{on}^{(m)}R^{(m)}\phi^{(m-1)}_--\lambda\phi^{(m)}_-. \label{eq phi minus m} 
\end{equation}

\begin{figure}[tp]
\begin{center}
\includegraphics[scale=1]{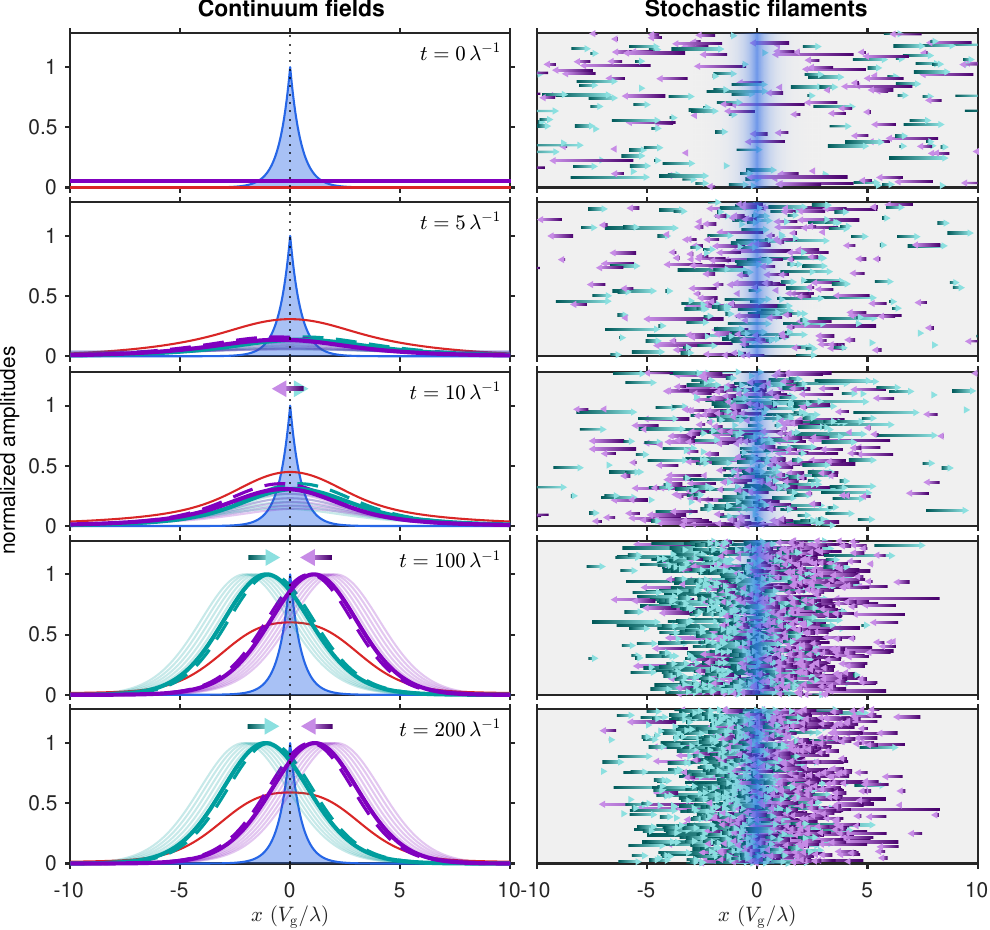}
\end{center}
\caption{{\bf Relaxation to steady state in the multistep model ($M=5$).}
Left column: time evolution of the continuum fields: Ran source $s(x)$ (blue, fixed), activated nucleator $c(x,t)$ (red), tubulin densities $\phi^{(0)}_\pm(x,t)$ (solid teal/purple), and plus-end densities $\eta_\pm(x,t)$ (dashed teal/purple), normalized by their steady-state peak values. Faint teal/purple curves show the similarly normalized intermediate densities $\phi^{(m)}_\pm(x,t)$ ($m=1,\ldots,M$), which for increasing $m$ are progressively separated from $\phi^{(0)}_\pm(x,t)$. Snapshots correspond to times $t = \{0, 5, 10, 100, 200\}\lambda^{-1}$.
Arrows at the top of the graphs indicate the polarity of the respective MT population; their arrowheads (plus ends) align with the peak of the plus-end densities $\eta_\pm$, and their lengths equal $\ell = V_\text{g}/\lambda$ for reference.
Right column: corresponding stochastic filament snapshots, with right-pointing MTs in teal and left-pointing MTs in purple.
Parameters: $M=5$, $V_\parallel/V_\text{g}=0.2$, with all other parameters as in Table~\ref{table params}. The system is initialized with a uniform seed of left/right MT populations and no active nucleators. In contrast to the single-step case at the same velocity ratio (Fig.~\ref{fig:dynamics}), the condensate relaxes to a correctly polarized, plus-end inward steady state.
}
\label{fig:dynamics-M5}
\end{figure}

Taken together, Eqs.~\eqref{eq phi plus m},~\eqref{eq phi minus m},~\eqref{eq phi pm 0},~\eqref{eq eta pm}, and~\eqref{eq: c approx M}, with the nucleation rate given by Eq.~\eqref{eq n m step}, form a closed, \emph{length-independent} dynamical system for the $2(M+2)+1$ fields: $\phi^{(m)}_\pm(x,t)$ ($0\leq m\leq M$), $\eta_\pm(x,t)$, and $c(x,t)$.

\subsection{Proper polarity sorting can emerge with measured velocities}
\label{sec:multistep-success}

We apply the same numerical scheme (Appendix~\ref{app:numerical-scheme}) to the multistep dynamical system, keeping $V_\parallel/V_\text{g} = 1/5$ fixed, and starting from the same uniform initial condition as in the single-step simulations. Figure~\ref{fig:dynamics-M5} (and Movie~\ref{mov:M5-eps02}) shows the resulting transient relaxation for $M=5$, a multistep analog of Fig.~\ref{fig:dynamics}. As anticipated, the network in this example settles into a correctly polarized, plus-end inward steady state, rather than the inverted one reached with $M=0$ at this velocity ratio. 

\begin{figure}[tp]
\begin{center}
\includegraphics[scale=1]{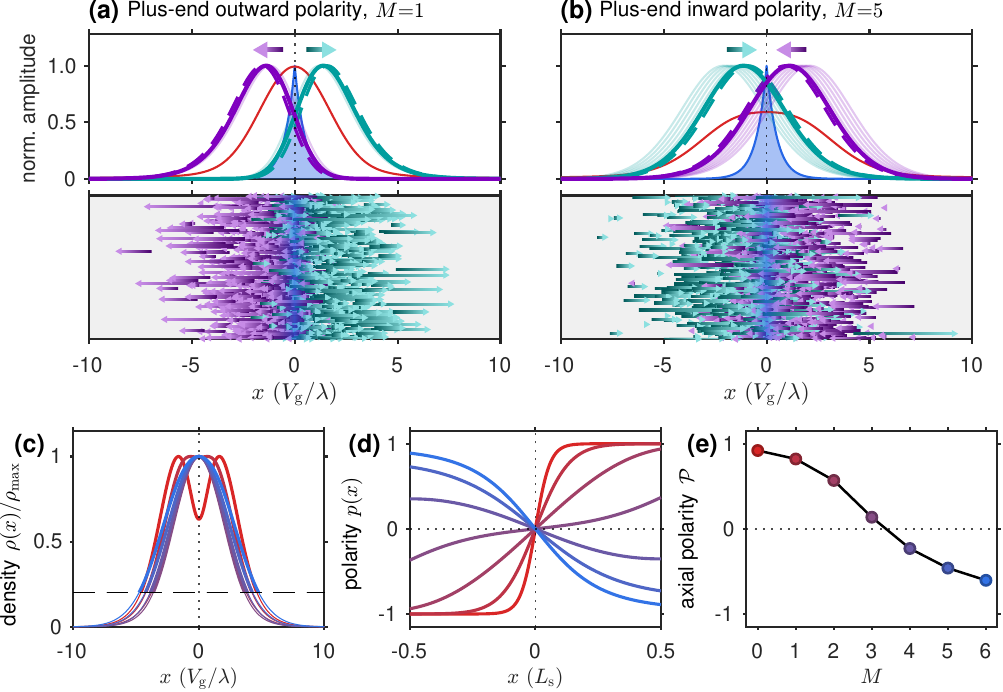}
\end{center}
\caption{{\bf The multistep model achieves correct spindle polarity at the measured velocity ratio.}
All panels show steady-state solutions (snapshots at $t=200\lambda^{-1}$) with $V_\parallel/V_\text{g}=0.2$ fixed.
\textbf{(a)} $M=1$ intermediate assembly stage: the system reaches plus-end outward (wrong) polarity.
\textbf{(b)} $M=5$ intermediate stages: the system reaches plus-end inward (correct) polarity.
Top row: Ran source $s(x)$ (blue, filled) and steady-state profiles of the Ran-activated species $c(x)$ (red), MT lattice densities $\phi^{(0)}_\pm$ (solid teal/purple), and plus-end densities $\eta_\pm$ (dashed). Faint teal/purple curves show the intermediate complex densities $\phi^{(m)}_\pm$ ($m=1,\ldots,M$), which for increasing $m$ are progressively separated from $\phi^{(0)}_\pm$.
Arrows at the top of the graphs indicate the polarity of the respective MT population; their arrowheads (plus ends) align with the peak of the plus-end densities $\eta_\pm$, and their lengths equal $\ell = V_\text{g}/\lambda$ for reference.
Bottom row: individual filament realizations (sampled from the steady-state $\psi_\pm(x,l)$; see Appendix~\ref{app:characteristics}), colored by MT polarity (teal: right-pointing; purple: left-pointing), with the Ran source density overlaid as a semi-transparent blue haze.
\textbf{(c)} Normalized MT density $\rho(x)/\rho_\text{max}$, with $\rho = \phi^{(0)}_+ + \phi^{(0)}_-$, for $M \in \{0,\ldots,6\}$ (red to blue); thick portion marks the spindle interior ($\rho > 0.2\,\rho_\text{max}$).
\textbf{(d)} Local polarity $p(x) = (\phi^{(0)}_+-\phi^{(0)}_-)/\rho$ within the spindle interior, $x$ normalized by spindle length $L_\text{s}$.
\textbf{(e)} Axial polarity $\mathcal{P}$ vs.\ $M$; $\mathcal{P}>0$ indicates wrong polarity (plus ends outward), $\mathcal{P}<0$ correct polarity (plus ends inward).
}
\label{fig:Mstep-fields}
\end{figure}

In Fig.~\ref{fig:Mstep-fields} we apply the axial polarity characterization of steady states, as done in Fig.~\ref{fig:M0-fields}, but with $M$ the varied control parameter in place of $V_\parallel/V_\text{g}$. When varying $M$, we rescale the binding affinity $\alpha$ to compensate for the reduced abundance of mature complexes at larger $M$ (Appendix~\ref{app:numerical-scheme}). Panels~\ref{fig:Mstep-fields}a,b show steady-state examples obtained for $M=1$ (see also Movie~\ref{mov:M1-eps02}) and $M=5$ (Fig.~\ref{fig:dynamics-M5}; Movie~\ref{mov:M5-eps02}), respectively. Panels~\ref{fig:Mstep-fields}c--e show density profiles, polarity profiles, and the axial polarity $\mathcal{P}$ across a range of $M$ values. For $M \leq 3$, the system reaches a steady state in which right-pointing (left-pointing) filaments accumulate predominantly on the right (left) half of the spindle (e.g., Fig.~\ref{fig:Mstep-fields}a), an outward polarity configuration akin to that of the single-step model (Fig.~\ref{fig:M0-fields}a). For $M \geq 4$, the picture reverses: right-pointing (left-pointing) filaments concentrate on the \emph{left} (right) half of the spindle (e.g., Fig.~\ref{fig:Mstep-fields}b), consistent with the minus-end outward polarity observed experimentally. The sign transition of $\mathcal{P}$ in Fig.~\ref{fig:Mstep-fields}e agrees quantitatively with the drift condition derived in Section~\ref{sec:drift}: intergenerational drift is directed toward the minus end when $V_\parallel > V_\text{g}/(M+2)$, first satisfied at $M = 4$ for $V_\parallel/V_\text{g} = 1/5$. The mechanism is intuitive: at this threshold, the polynomial distribution of mature complexes (integrand in Eq.~\eqref{eq phi m plus}) is sufficiently biased toward the minus end that, across branching generations, the network progressively adopts the correct polarity.

\begin{figure}[tp]
\begin{center}
\includegraphics[scale=1]{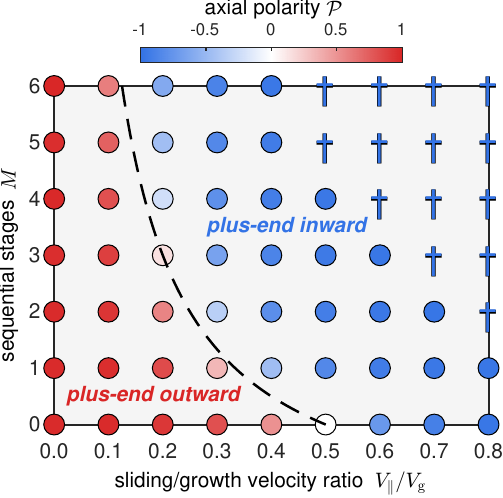}
\end{center}
\caption{{\bf State diagram of spindle polarity in the multistep model.}
Steady-state axial polarity $\mathcal{P}$ (color gradient) as a function of the velocity ratio $V_\parallel/V_\text{g}$ and the number of sequential assembly stages $M$. Red: plus-end outward (incorrect) polarity; blue: plus-end inward (correct) polarity. Dashed curve: analytical boundary $M+2 = V_\text{g}/V_\parallel$ (Section~\ref{sec:drift}). Dagger symbols ($\dagger$): transient condensates slide fast and die out (no steady-state structure).}
\label{fig:phase-diagram}
\end{figure}

Figure~\ref{fig:phase-diagram} extends this comparison to the full parameter space spanned by $V_\parallel/V_\text{g}$ and $M$. The axial polarity $\mathcal{P}$ transitions from positive (red; inverted polarity) to negative (blue; correct polarity) as either parameter increases. The analytical boundary $M = V_\text{g}/V_\parallel - 2$ (dashed curve) predicts the inversion line across the entire plane, supporting our single mother-daughter drift argument (Section~\ref{sec:drift}) as the organizing principle of spindle polarity. We note that the numerical transition appears slightly shifted relative to the analytical curve for two reasons. First, $\mathcal{P}$ is evaluated with respect to $\phi_\pm^{(0)}$, which is most directly accessible in fluorescence microscopy, rather than $\phi_\pm^{(M)}$, on which the analytical argument hinges. Second, that argument treats the Ran-activated nucleator as a spatially uniform reservoir, whereas in our simulations $c(x)$ is peaked around the chromosomes, which perturbs the branchpoint drift (Appendix~\ref{app:gradient-drift}).

\subsection{Branching statistics as a readout of the number of assembly steps}
\label{sec:mean-branching-time}

The drift argument of Section~\ref{sec:drift} shows that correct polarity requires a sufficient number of rate-limiting assembly steps, but the effective $M$ is not known experimentally. Global polarity is only an indirect readout of $M$, as it also depends on the velocity ratio and on the nucleator profile. We therefore ask which measurable features of multistep branching depend more directly on $M$. Because mature complexes take time to assemble, $M$ sets both \emph{when} a mother MT branches (that is, its age at branching) and \emph{where along its length} the typical branch forms. These single-filament statistics do not depend on the sliding speed or on the spatial organization of the population, and so provide more direct experimental readouts of $M$.

At steady state, spatially integrated nucleation balances catastrophe, $\int n_\pm\,dx = \lambda N_\pm$, with $N_\pm = \int\eta_\pm\,dx$ the number of MTs of each polarity. Each MT is therefore replaced, on average, by a single daughter over its lifetime, although individual mothers may produce no daughters or several. Below, we consider the distribution of parent ages sampled across branching events in this steady-state population; mothers producing several daughters contribute several events. During population growth, both nucleator availability and the population's age distribution evolve, so the steady-state branching statistics derived below need not apply.

\begin{figure}[tp]
\begin{center}
\includegraphics[scale=1]{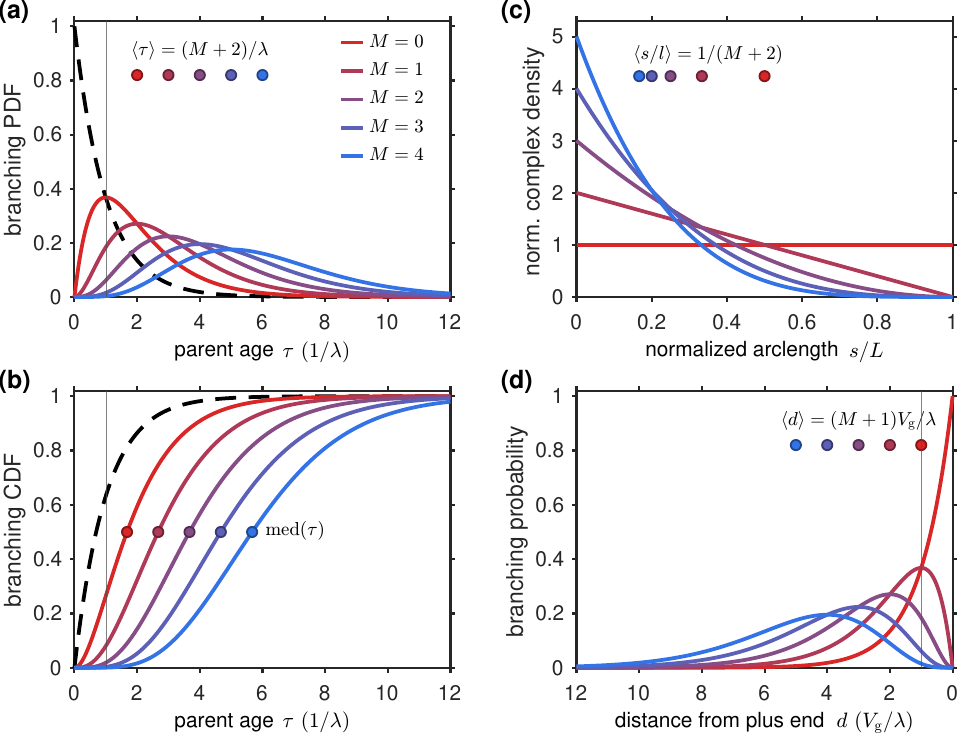}
\end{center}
\caption{{\bf Branching statistics in the multistep model.} Colored curves span $M=0$ to $M=4$ along a red-to-blue gradient. Filled markers in (a), (c), and (d) mark the distribution means; those in (b) mark the medians (CDF${}=0.5$).
\textbf{(a)} Parent-age distribution at branching: the probability density that a branching event occurs on a mother MT of age $\tau$. It is a Gamma distribution of shape $M+2$ and rate $\lambda$ (Eq.~\eqref{eq:gamma-distribution}), with mean $(M+2)/\lambda$. The dashed line is the MT survival factor $e^{-\lambda\tau}$, shown for reference.
\textbf{(b)} Cumulative distribution functions of the age distributions in (a). At the typical MT lifetime $\tau_\text{MT}=1/\lambda$, the CDF decreases substantially with increasing $M$: a progressively smaller fraction of branching events has occurred on average-lived MTs. The dashed line is $1-e^{-\lambda\tau}$, the exponential lifetime CDF.
\textbf{(c)} Normalized density of mature branching complexes as a function of normalized arclength $s/L$ along the mother MT (integrand of Eq.~\eqref{eq phi m plus}), with $s/L=0$ the minus end and $s/L=1$ the plus end. The distribution is uniform for $M=0$ and increasingly biased toward the minus end as $M$ grows (linear for $M=1$, quadratic for $M=2$, cubic for $M=3$, and so on), with mean position $1/(M+2)$ from the minus end.
\textbf{(d)} Population-level probability density that branching occurs at absolute distance $d$ from the plus end (in units of $\ell=V_\text{g}/\lambda$), obtained by accounting for the steady-state distribution of MT lengths: a Gamma distribution of shape $M+1$, $\propto d^{M} e^{-d/\ell}$, with mean $(M+1)\ell$. The horizontal axis is reversed, placing the plus end on the right.}
\label{Fig multistep statistics}
\end{figure}

We define $\tau_n$ as the mean parent age sampled across branching events. Assuming that the unbound nucleator concentration $c_\text{u}$ is uniform over the length of a parent MT, its branching rate is proportional to the total number of mature complexes it hosts. According to Eq.~\eqref{eq phi m plus}, this quantity, integrated along the MT length, scales as $l^{M+1}\propto \tau^{M+1}$, with $\tau$ the age of the parent. Accounting for MT turnover at rate $\lambda$, the likelihood of branching must also be weighted by the survival factor $e^{-\lambda \tau}$. The resulting expression ($\sim\tau^{M+1}e^{-\lambda \tau}$) is proportional to a Gamma distribution with shape parameter $M+2$ and rate parameter $\lambda$ (Fig.~\ref{Fig multistep statistics}a):
\begin{equation}
    f(\tau;\,M{+}2,\lambda)=\frac{\lambda^{M+2}}{\Gamma(M+2)}\tau^{M+1}e^{-\lambda\tau}. \label{eq:gamma-distribution}
\end{equation}
The average nucleation time $\tau_n$ is then defined as its expectation value:
\begin{equation}
\tau_n := \langle\tau\rangle
= \frac{M+2}{\lambda}.
\end{equation}
Hence, the typical time elapsed between the birth of a parent MT and the nucleation of a new ``daughter'' MT along it is proportional to $M+2$ and inversely proportional to the catastrophe rate $\lambda$; the corresponding filament length is $l=(M+2)V_\text{g}/\lambda$, a factor of $(M+2)$ longer than the typical MT length. The buildup of branching complexes through a sequence of rate-limiting steps introduces a delay that biases branching toward older, longer MTs, at the tail of the age (or length) distribution. For the single-step model ($M=0$), one has $\tau_n=2/\lambda$, which is already twice the mean MT lifetime. This follows from the fact that the number of lattice sites offered for potential branching events increases linearly in time as the parent MT grows. 

The cumulative distribution function (CDF) of the branch time probability density function (PDF) gives the fraction of branching events occurring on parents younger than a given age. Its analytical form, given in Appendix~\ref{app:branching-probability} and plotted in Fig.~\ref{Fig multistep statistics}b, shows that the share of branching events on long-lived MTs (with $\tau>1/\lambda$) becomes increasingly extreme as $M$ grows. Hence, at steady state, filaments that survive well beyond $1/\lambda$ by chance account for a disproportionate fraction of all branching events. At steady state, branching events are therefore concentrated on a minority of long-lived MTs, but this does not make the population fragile: when MTs are scarce, nucleators are shared among fewer complexes, so that MTs of all ages branch more readily.

In Fig.~\ref{Fig multistep statistics}c we plot, as a function of the normalized arclength coordinate $s/L$, with $L$ the length of the mother filament, the kernel of Eq.~\eqref{eq phi m plus} that sets the density of mature ($M$-staged) branching complexes: a degree-$M$ polynomial that, for $M\geq1$, vanishes at the plus end ($s/L=1$) and rises to its maximum at the minus end ($s/L=0$). In Fig.~\ref{Fig multistep statistics}d we show the resulting population-level distribution of branch positions, as a function of the \emph{absolute} distance $d$ from the plus end (in units of $\ell=V_\text{g}/\lambda$). A mature complex can sit at distance $d$ only on a mother of length at least $d$, where it contributes the kernel value $(l-s)^M=d^M$, independent of the mother length $l$. Summing over the population then gives
\begin{equation}
P(d) \propto \int_{d}^{\infty} d^M\, e^{-l/\ell}\,dl \propto  d^M e^{-d/\ell}, \label{eq:Pd}
\end{equation}
where $d^M$ pulls out of the integral and the remaining factor $\int_d^\infty e^{-l/\ell}\,dl\propto e^{-d/\ell}$ is the probability that the mother reaches length $d$. This is a Gamma distribution of shape parameter $M+1$, whose mean $\langle d\rangle = (M+1)\,\ell$ shifts away from the plus end, toward the minus end, as $M$ grows. Panels~\ref{Fig multistep statistics}c and~\ref{Fig multistep statistics}d correspond to the branch-nucleation quantifications of Thawani et al.~\cite{thawani2019spatiotemporal} (their Figs.~2B and~2A, respectively), with one difference in sampling: they recorded the first branching event on each initially bare mother during network growth, whereas our distributions include all branching events in a steady-state population. The fractional position (Fig.~\ref{Fig multistep statistics}c) is insensitive to this difference, because along a mother of given length the density of mature complexes has the same shape for its first and later branching events.
It is worth noting that this $M$-sensitivity is a special feature of measuring branch positions from the plus end. Repeating the construction from the \emph{minus end}, at absolute arclength $s$, the local density is the kernel $(l-s)^M$ itself, which now depends on the mother length $l$; summing over the population then gives
\begin{equation}
    P(s) \propto \int_{s}^{\infty} (l-s)^M\, e^{-l/\ell}\,dl
 \overset{w=l-s}{=} e^{-s/\ell}\!\int_0^\infty\! w^M e^{-w/\ell}\,dw
 \propto e^{-s/\ell}. \label{eq:Ps}
\end{equation}
The $M$-dependence disappears upon normalization: measured as a function of absolute distance from the minus end, branch positions are exponentially distributed with mean $\ell=V_\text{g}/\lambda$. Hence, the informative, $M$-sensitive readout is $P(d)$ of Fig.~\ref{Fig multistep statistics}d and not $P(s)$.
\section{A unified view of assembly-dependent polarity sorting}
\label{sec:generalization}

The mechanism identified above does not hinge on the number of sequential assembly steps per se, but rather on the delay that assembly introduces (Section~\ref{sec:mean-branching-time}): mature branching complexes accumulate on lattice sites of long residence time, which have drifted backward with the mother MT. 
To emphasize this point, we consider an alternative nucleation model in which a delay is caused by cooperative association rather than sequential steps (Appendix~\ref{app:cooperative}). With one intermediate species ($M=1$) that recruits further copies of itself at a self-recruitment rate $a_\text{s}$, this cooperative model can produce the requisite minus-end bias and thereby yield inward polarity at the measured velocity ratio (Fig.~\ref{fig:cooperative-fields}). Here we derive a general criterion for proper polarity sorting in the presence of branching nucleation that is applicable to the single-step, multistep, and cooperative models, as well as any other model in which the branching rate depends on the residence time of lattice sites.

We denote by $u=(l-s)/V_\text{g}$ the residence time of a lattice site, i.e., the time elapsed since its incorporation at the plus end. Assuming a spatially uniform, time-independent available nucleator concentration $c_\text{u}$, the branching rate at a lattice site depends on $u$ alone, through a nonnegative function $K(u)$. The single-step model corresponds to a constant $K$, the multistep model to $K\propto u^M$ (Eq.~\eqref{eq phi m plus}), and the cooperative model to $K\propto e^{a_\text{s}u}-1$ (Eq.~\eqref{eq phi1 self def}). As in Section~\ref{sec:mean-branching-time}, we sample branching events across the steady-state population. For a mother MT nucleated at $x_0$, the lattice site at distance $s$ from its minus end at age $\tau$ has $u=\tau-s/V_\text{g}$. Weighting by the survival factor, the density of branching events is proportional to $e^{-\lambda\tau}K(\tau-s/V_\text{g})$, for $0\leq s\leq V_\text{g}\tau$. In terms of $s$ and $u$, it factorizes:
\begin{equation}
    p(s,u)\propto e^{-\lambda s/V_\text{g}}\,e^{-\lambda u}K(u),\qquad s,u\geq0.
    \label{eq:factorized-branching}
\end{equation}
We restrict attention to kernels for which $e^{-\lambda u}K(u)$ has a finite, nonzero integral and a finite first moment; for the cooperative model, this requires $a_\text{s}<\lambda$. For any such $K$, the branch distance $s$ from the minus end is exponentially distributed with mean $\ell=V_\text{g}/\lambda$, generalizing $P(s)$ of Section~\ref{sec:mean-branching-time} (Eq.~\eqref{eq:Ps}) and the $a_\text{s}$-independent result $\langle \bar{s} \rangle=\ell$ of Appendix~\ref{app:cooperative-drift}. The residence time at branching has probability density proportional to $e^{-\lambda u}K(u)$ which, with $d=V_\text{g}u$, gives the distribution of branch distances from the plus end, illustrated for the multistep model in Eq.~\eqref{eq:Pd} and Fig.~\ref{Fig multistep statistics}d. The relevant delay is therefore the lattice residence time sampled across branching events: multiplying $K(u)$ by a constant changes the overall branching rate but leaves this normalized distribution unchanged.

The daughter MT is nucleated at the displacement $\Delta=s-V_\parallel\tau=(1-V_\parallel/V_\text{g})\,s-V_\parallel u$ from $x_0$. The first term is the forward advance of the mother's plus end at the time the branching site was incorporated, and the second is the backward sliding of that site since then. Averaging with Eq.~\eqref{eq:factorized-branching},
\begin{equation}
    \langle\Delta\rangle=\frac{V_\text{g}-V_\parallel}{\lambda}-V_\parallel\langle u\rangle,
    \label{eq:general-displacement}
\end{equation}
so the intergenerational drift is directed toward the minus end if and only if
\begin{equation}
    \lambda\langle u\rangle>\frac{V_\text{g}-V_\parallel}{V_\parallel}
    \quad\Longleftrightarrow\quad
    \langle d\rangle>\frac{V_\text{g}-V_\parallel}{V_\parallel}\,\ell.
    \label{eq:general-criterion}
\end{equation}
For the single-step, multistep, and cooperative models, $\lambda\langle u\rangle=1$, $M+1$, and $(2\lambda-a_\text{s})/(\lambda-a_\text{s})$, recovering $V_\parallel>V_\text{g}/2$, $V_\parallel>V_\text{g}/(M+2)$ and Eq.~\eqref{eq:cooperative-drift-condition}, respectively.

Equation~\eqref{eq:general-criterion} gives a general criterion for backward intergenerational drift, the mechanism underlying polarity sorting in the models studied here. The direction of drift is set by the mean residence time of branching sites in units of $1/\lambda$, compared with the ratio of the plus-end velocity $V_\text{g}-V_\parallel$ to the sliding speed $V_\parallel$. At the reference ratio $V_\parallel/V_\text{g}=1/5$, the mean residence time at branching must exceed four mean MT lifetimes, or equivalently the mean branch distance from the plus end must exceed $4\ell$. The criterion is independent of how the delay is produced and involves only measurable quantities: the distribution of branch positions relative to the plus end \cite{thawani2019spatiotemporal} (sampled over all branching events), the mean MT length $\ell$, and the sliding and growth velocities. It assumes steady-state sampling, spatially uniform and time-independent nucleator availability $c_\text{u}$ (see Appendix~\ref{app:gradient-drift} for the treatment of gradients), exponentially distributed MT lifetimes, constant growth and sliding speeds, and strictly parallel nucleation. Under these assumptions, the above criterion determines the sign of the mean intergenerational displacement, which, we argue, largely determines the transition in the spindle's axial polarity.

\section{Discussion}
\label{sec:discussion}

We have shown that, with $V_\parallel/V_\text{g} < 1/2$ being the experimentally measured ratio of sliding/growth velocities, accounting for the sequential, multistep nature of the branching process is sufficient to reproduce the observed minus-end outward spindle polarity. The underlying mechanism combines motor-driven transport with the kinetics of branching: (1) the minus ends of the growing `mother' filaments persistently slide backward, despite rapid growth at the plus ends; and (2) under rate-limiting sequential assembly, mature branching complexes accumulate near those minus ends, where the tubulin lattice is oldest (Eq.~\eqref{eq phi m plus}, Fig.~\ref{Fig multistep schematic}). Taken together, new daughter filaments, which branch from these complexes and inherit the mother's polarity, are nucleated backward relative to the site at which the mother was first nucleated. This enables a minus-end–directed intergenerational drift, in which the self-renewing MT population effectively moves backward even though the center of mass of each individual MT moves forward by plus-end growth. For a uniform background of nucleators, as holds approximately in the middle of the spindle, the intergenerational drift is negative when $V_\parallel/V_\text{g} > 1/(M + 2)$ (Section~\ref{sec:drift}), a condition that, at the velocity ratio measured in \textit{Xenopus} egg extract spindles, is satisfied only for $M \geq 4$. At smaller $M$, the bias is insufficient to counteract plus-end polymerization, and the system relaxes to an inverted, plus-end outward polarity (Section~\ref{sec:multistep-success}).

This sharp dependence on $M$ provides a direct quantitative link between the molecular kinetics of MT branching nucleation and polarity sorting in the spindle. Establishing such a link was precisely the prospect raised by Thawani et al.~\cite{thawani2019spatiotemporal} at the close of their study: that a quantitative understanding of how molecular effectors organize autocatalytic nucleation in space and time would illuminate how the spindle is constructed through spatially regulated MT generation and motor-based transport. Our integrated model shows that, with the measured motor-driven transport (slow relative to the growth speed), the specific form of the branching reaction, namely the number of rate-limiting assembly steps $M$, is a critical parameter determining whether the MT network as a whole organizes into the observed, inward-pointing polarity configuration.

The requirement $M \geq 4$ is plausible given the proposed architecture of the branching complex. Experiments in \textit{Xenopus} egg extracts and with purified \textit{Xenopus} proteins have suggested at least three sequential steps: TPX2 first associates with the parent MT \cite{thawani2019spatiotemporal}, then recruits the augmin complex \cite{alfaro2020biochemical}, which in turn recruits $\gamma$-TuRC to template the new filament \cite{song2018mechanism,alfaro2020biochemical}.
A natural candidate for a fourth rate-limiting step is nucleation from $\gamma$-TuRC itself, whose rate-limiting step is the cooperative assembly of several $\alpha\beta$-tubulin dimers on the ring \cite{thawani2020transition,consolati2020microtubule,wieczorek2021biochemical} (approximately four for \textit{Xenopus} $\gamma$-TuRC \cite{thawani2020transition}, and possibly more for human $\gamma$-TuRC \cite{consolati2020microtubule}), accompanied by a partial or complete closure of $\gamma$-TuRC \cite{vermeulen2024asymmetry,brito2024transition,aher2024structure}. Whether these molecular events provide four sequential stages that are each slow relative to MT turnover remains to be established experimentally.

Beyond the step count, we show in Appendix~\ref{app:cooperative} that cooperative self-recruitment of branching factors offers a complementary route to achieving the requisite minus-end bias for inward-pointing polarity. In the spindle, both the multistep and cooperative pathways may well act in concert. A more speculative alternative, not modeled here and not based strictly on an assembly delay, is that minus-end–directed motors actively transport nucleation factors as cargo toward the minus ends of mother MTs before those factors are incorporated into mature branching complexes. Branching hubs would then migrate toward the minus end faster than the MT lattice itself, accelerating the required rearward bias through active transport rather than through the residence-time mechanism studied here. We leave its investigation to future work.

The present study is confined to one spatial dimension with two antiparallel filament orientations, corresponding to the fully nematic limit. It therefore does not address how nematic order first arises, the emergence of a spindle interface, or overall spindle shape. Addressing such aspects would require extending our framework to higher spatial (and orientational) dimensions. 

Dynein motors, which are known to focus the spindle poles \cite{heald1996self,gaetz2004dynein,mitchison2005roles,dalton2022gelation}, are absent from the current model. Their inclusion is expected to produce a nonuniform sliding speed: decreased near the poles and perhaps slightly increased around the midzone \cite{burbank2007slide}, modulations that would feed into the intergenerational drift discussed here. Dynein may also contribute to polarity sorting more directly: filament-based simulations without branching nucleation suggest that dynein-like motors could, for a suitable binding mechanism, establish inward polarity even when sliding is much slower than growth \cite{chew2025molecular}. We emphasize, however, that inward-pointing polarity is obtained experimentally with \emph{or without} active dynein motors \cite{burbank2007slide,dalton2022gelation}. We have also neglected rescues, which are rare in \textit{Xenopus} egg extract spindles. Since a rescue preserves old lattice, and longer MTs are more likely to be rescued, we would expect rescues, if anything, to strengthen the bias toward old branching sites; a quantitative treatment is beyond the scope of this work. 

We have further made two assumptions about the Ran-activated species, of concentration $c(x,t)$, which localizes branching around chromosomes: that it binds and unbinds rapidly, and that it acts as the final catalytic step of the assembly chain. Either assumption could be relaxed; $c$ could be treated as a slow component and/or as an earlier step in the chain. Doing so, however, would break the exact kernel representation (Eqs.~\eqref{eq phi m plus},~\eqref{eq phi m minus}), which requires every assembly intermediate to be drawn from a spatially uniform reservoir, which $c$ is crucially not. Finally, we have verified that the free parameters governing the Ran-activated species ($\ell_s$, $s_0$, $a_c$, $\lambda_c$, and $\alpha$) do not affect our central conclusions. We tried a wide range of values for each: they reshape the $c$ profile (as expected), and thereby the spatial extent and amplitude of the model spindle, but leave unchanged where the transition from outward to inward polarity first occurs. The analytical crossover criterion $V_\parallel/V_\text{g} > 1/(M+2)$ (Section~\ref{sec:drift}) is robust provided that $c$ is nearly flat across the midzone.

From a mathematical perspective, a convenient and nontrivial property of our modeling framework is that filament length is exactly marginalized as an independent variable, reducing the full kinetic problem (which \textit{a priori} spans position, length, and time) to a closed system in space and time alone, without approximation, and for arbitrary $M$ (Section~\ref{sec:reduced-dim-multistep}). This reduction follows directly from the structure of the growth and nucleation terms: integrating the kinetic equation over length couples each length moment of the population only to the one below it, forming a finite chain that ends at the plus-end density, which is in turn fed by the nucleation rate. Because the coupling runs strictly downward, the chain closes on itself, with no closure approximation. In this sense it differs from the \emph{orientational} order parameters (or moments) of Doi--Edwards theory, which couple to ever-higher ones in the hierarchy and can therefore only be closed approximately. For a three-dimensional spindle with two orientational degrees of freedom, the full kinetic description would be a $(6+1)$-dimensional problem; the length marginalization reduces it to $(5+1)$ dimensions, making it potentially tractable under an orientational moment closure. At any rate, an axisymmetric model would reduce to a tractable $(3+1)$-dimensional problem. 

Beyond the spindle, the framework developed here could apply to other autocatalytic filament networks in which branching requires the sequential, rate-limiting assembly of molecular factors along a growing mother filament. Branched actin networks, nucleated via the Arp2/3 complex in lamellipodia and other cellular protrusions, are natural candidates. There, branching involves activation of Arp2/3 by nucleation-promoting factors (NPFs such as WASp, N-WASP, or WAVE) and association with an existing actin filament, through multiple binding steps \cite{campellone2010nucleator,blanchoin2014actin}. If these steps support sequential, rate-limiting accumulation of intermediates on the mother lattice, the same residence-time argument could apply: mature branching complexes would accumulate toward the pointed end of the mother filament, where actin subunits are oldest. Under these assumptions, newly formed Arp2/3-mediated branches would be enriched there, increasingly so with the number of rate-limiting assembly steps $M$. Their spatial distribution could therefore help constrain the effective number of rate-limiting steps, provided that spatial variations in factor availability and subsequent branch loss are accounted for. The mathematical framework, including the exact length marginalization, could carry over under analogous growth and assembly assumptions, with the strictly parallel branching assumption replaced by the characteristic $\sim70^\circ$ branching angle of Arp2/3~\cite{mullins1998interaction}, which is substantially larger than the near-parallel branching of augmin–$\gamma$-TuRC–mediated MT nucleation.

\section*{Code availability}
The MATLAB code used for the simulations and figure generation
is publicly available at
\url{https://github.com/ido-lavi/spindle1D/}.

\section*{Acknowledgments}
The Flatiron Institute is a division of the Simons Foundation. S.F. acknowledges funding from the Vienna Science and Technology Fund (WWTF) [10.47379/VRG20002] and the Volkswagen Foundation (Az. 0201568-01).

\newpage
\begin{appendices}
\section{Force balance on filaments}
\label{app:force-balance}
Here, we discuss the microscopic force balance leading to the MT minus-end velocities $\mp V_\parallel$ appearing in Eq.~\eqref{eq kinetic growth}. Our starting point is an explicit description of highly crosslinked, rigid filaments, with $x_i$, $l_i$, and $p_i=\pm1$ denoting the minus-end position, length, and polarity of filament $i$. The force exerted on this filament by crosslinkers that connect it to a nearby filament $j$ is given by
\begin{equation}
F_{ij}=\int_0^{l_i}ds_i\int_0^{l_j}ds_j\int_{\Omega(x_i+s_ip_i)}\delta(y-x_j-s_jp_j)f_{ij}\,dy,
\end{equation}
where $s_i,s_j$ are arclength coordinates on the respective filaments and $\Omega(x_i+s_ip_i)$ is defined as the region within which crosslinkers, with one head bound to position $s_i$ along filament $i$, can extend (see Fig.~\ref{Fig schematic basic}). The interaction force density, satisfying $f_{ij}=-f_{ji}$, is defined by
\begin{equation}
    f_{ij}=-\gamma\left(\dot{x}_i-\dot{x}_j+V_\parallel (p_i-p_j)\right).\end{equation}
Here, $\gamma$ denotes an effective drag arising from the external stretching of crosslinkers, and $V_\parallel$ is the effective speed at which each motor head walks toward the plus end of the MT
 to which it is bound. The interfilament friction and active terms in this force density have been derived from crosslinker thermodynamics for filament pairs~\cite{barutel2026entropic}. This force density assumes a uniform arrangement of multimeric motors (vertebrate kinesin-5, also known as Eg5), which generate active sliding forces between antiparallel filaments (i.e., $p_j = -p_i$; see Fig.~\ref{Fig schematic basic}). Such behavior has been demonstrated \emph{in vitro} \cite{kapitein2005bipolar}, and these motors have been shown to drive poleward MT flux in \emph{Xenopus} egg extract spindles \cite{miyamoto2004kinesin}.
 We note that, in the general theory, $f_{ij}$ may accommodate multiple crosslinker types, including those with asymmetric head distributions along filaments, and in three dimensions \cite{furthauer2021design}.

In the overdamped limit, and further assuming that drag with the surrounding medium is negligible compared with crosslink forces~\cite{furthauer2021design,barutel2026entropic}, the force-balance equation on filament $i$ reads
\begin{equation}
    \sum_{j}F_{ij}=0. \label{eq force balance i}
\end{equation}
This condition is trivially satisfied when $f_{ij}=0$ for all neighbors $i,j$. Provided the crosslink network is connected, the solution simply reads $\dot{x}_i=-V_\parallel p_i+v_0$ where $v_0$ is a constant velocity of the reference frame. Imposing that $v_0=0$,
\begin{equation}
    \dot x_i=-V_\parallel p_i. \label{eq flux i}
\end{equation}
This key result states that all members of the highly crosslinked ensemble move backward relative to their individual polarity at a speed independent of their position $x$. Even if a filament does not directly interact with an antiparallel neighbor, so that the sliding force on it vanishes, it still follows the same pattern of motion, because it experiences drag from parallel neighbors that, elsewhere in the network, are explicitly moved against the antiparallel population (see Fig.~\ref{Fig schematic basic}). This double gel-like behavior of the ensemble is consistent with observations in \textit{Xenopus} egg extract spindles at sufficiently high crosslinker density \cite{dalton2022gelation} and further supported by particle-based simulations \cite{dalton2022gelation}. 

\section{Solution by the method of characteristics}
\label{app:characteristics}

\subsection{Generic solution}
Equation~\eqref{eq kinetic growth} is a first-order linear PDE; its characteristics are the trajectories of individual filaments in $(x,l,t)$ space. A filament nucleated at position $x_0$ and time $t_0$ has, at later time $t$, age $\tau(t) = t - t_0$, length $l(t) = V_\text{g}\tau$, and minus-end position $x(t) = x_0 \mp V_\parallel\tau$ for the ($\pm$) populations, respectively. Along each such trajectory, Eq.~\eqref{eq kinetic growth} reduces to the ordinary differential equation 
\begin{equation}
\tfrac{d}{dt}\psi_\pm(x(t),l(t),t) = -\lambda \psi_\pm(x(t),l(t),t),
\end{equation}
and therefore,
\begin{equation}
    \psi_\pm(x,l,t)=\psi_\pm(x_0,0,t_0)e^{-\lambda(t-t_0)}=V_\text{g}^{-1}n_\pm(x_0,t_0)e^{-\lambda(t-t_0)},
\end{equation}
where we used the boundary condition Eq.~\eqref{eq flux condition} to set the initial value on each characteristic. 
Expressing the birth coordinates $(x_0,t_0)$ in terms of the observation coordinates $(x,l,t)$ gives $t_0 = t - l/V_\text{g}$ and $x_0 = x \pm V_\parallel l/V_\text{g}$ respectively for the ($\pm$) populations. Hence,
\begin{equation}
\psi_\pm(x,l,t) = V_\text{g}^{-1}\, n_\pm\!\left(x \pm \frac{V_\parallel l}{V_\text{g}},\; t - \frac{l}{V_\text{g}}\right) e^{-\lambda l/V_\text{g}}.
\end{equation}
This formula exactly satisfies Eqs.~\eqref{eq kinetic growth} and~\eqref{eq flux condition} and holds for filaments born after $t = 0$ (i.e., those with $l < V_\text{g}t$). Filaments present at $t = 0$ contribute the first branch of the general solution Eq.~\eqref{eq:characteristics}, with initial data advected and decayed along the same linear characteristics.

\subsection{Steady-state reduction}
At steady state, the nucleation rate is time-independent: $n_\pm(x,t)=n_{\text{s},\pm}(x)$. For any fixed $l$, the initial-condition contribution $\psi_{\pm,0}(x \pm V_\parallel t,\, l - V_\text{g}t)\,e^{-\lambda t}$ is supported only for $l \geq V_\text{g}t$ and vanishes as $t \to \infty$ in any case. So only the self-sustaining autocatalytic population persists. The steady-state $\psi_{\text{s},\pm}(x,l)$ is therefore:
\begin{equation}
\psi_{\text{s},\pm}(x,l) = V_\text{g}^{-1} n_{\text{s},\pm}(x \pm V_\parallel l/V_\text{g})\, e^{-\lambda l/V_\text{g}}. \label{eq psi pm steady}
\end{equation}

Since $n_\pm$ is a functional of $\psi_\pm$, Eq.~\eqref{eq psi pm steady} is in general a self-consistency condition for any steady state. If $n_{\text{s},\pm}$ is known but $\psi_{\text{s},\pm}$ is not, as is the case in our reduced-dimension framework, we may use Eq.~\eqref{eq psi pm steady} to reconstruct the full number densities exactly.

Conversely, solving Eq.~\eqref{eq psi pm steady} for $n_{\text{s},\pm}$ gives the inverse relation:
\begin{equation}
n_{\text{s},\pm}(x) = V_\text{g}\,\psi_{\text{s},\pm}\!\left(x \mp \frac{V_\parallel l}{V_\text{g}},\, l\right) e^{\lambda l/V_\text{g}}, \qquad \forall\, l \geq 0. \label{eq n pm steady}
\end{equation}
At $l=0$ this is simply the boundary condition Eq.~\eqref{eq flux condition}: $n_{\text{s},\pm}(x)=V_\text{g}\,\psi_{\text{s},\pm}(x,0)$. For $l>0$, Eq.~\eqref{eq n pm steady} reads: the nucleation rate at $x$ equals $V_\text{g}$ times the density of filaments of length $l$ whose minus end is at $x \mp V_\parallel l/V_\text{g}$, corrected for the exponential decay over their age $l/V_\text{g}$. That this expression must be independent of $l$ is itself a nontrivial property of any steady-state solution.

\subsection{\texorpdfstring{Stochastic sampling of $\psi_{\text{s},\pm}$}{Stochastic sampling of psi\_s,pm}}
The reconstruction Eq.~\eqref{eq psi pm steady} allows us to draw individual filaments for visualization from the reduced-dimension solution, producing the filament snapshots that accompany the field profiles in Figs.~\ref{fig:M0-fields} and~\ref{fig:Mstep-fields}. Given the converged steady-state nucleation rate $n_{\text{s},\pm}(x)$, we first reconstruct the full density $\psi_{\text{s},\pm}(x,l)$. In dimensionless units ($V_\text{g}=\lambda=1$, $\varepsilon=V_\parallel/V_\text{g}$), Eq.~\eqref{eq psi pm steady} reads
\begin{equation}
    \psi_{\text{s},\pm}(x,l) = n_{\text{s},\pm}(x \pm \varepsilon l)\, e^{-l},
\end{equation}
with $x$ the minus-end position and $l$ the filament length. We evaluate this on a length grid $l\in[0,l_\text{max}]$ (with $l_\text{max}$ several times the mean MT length, beyond which the factor $e^{-l}$ makes the density negligible), applying the spatial shift $x\mapsto x\pm\varepsilon l$ spectrally by multiplying $\hat{n}_{\text{s},\pm}$ by $e^{\pm i k\varepsilon l}$ before the inverse transform. 

To sample filaments, we treat the grid values $\psi_{\text{s},\pm}(x_i,l_j)$ as an unnormalized joint probability mass over discrete (minus-end position, length) pairs: multiplied by the cell area $\Delta x\,\Delta l$, each value is the expected filament count in that cell. The number of filaments of each polarity is set by the steady-state balance of nucleation flux against turnover,
\begin{equation}
    N_\pm = \iint \psi_{\text{s},\pm}(x,l)\,dx\,dl = \int n_{\text{s},\pm}(x)\,dx,
\end{equation}
rounded up to an integer. We then draw $N_\pm$ index pairs independently with dense regions of $\psi_{\text{s},\pm}$ sampled in proportion to their weight. Each sampled pair $(x_i,l_j)$ becomes a rod: a right-pointing ($+$) teal filament with minus end at $x_i$ and plus end at $x_i+l_j$, or a left-pointing ($-$) purple filament with minus end at $x_i$ and plus end at $x_i-l_j$. The transverse coordinate $y_i$ of each rod is assigned uniformly at random solely to separate the filaments for display. 
Unlike the time-resolved stochastic simulations of Appendix~\ref{app:gillespie} (used for Figs.~\ref{fig:dynamics},~\ref{fig:dynamics-M5} and the supplementary movies), this procedure samples the \emph{steady state} directly, and its output agrees closely with the late snapshots of those simulations.

\section{Measured parameters of MT dynamics}
\label{app:measured-parameters}
The parameters characterizing MT dynamics, aside from those defining the nucleation model, are the catastrophe rate $\lambda$, the sliding speed $V_\parallel$, and the polymerization speed $V_\text{g}$. Experimental measurements of these quantities, summarized in Table~\ref{table measured}, provide a basis for selecting parameter values and testing the predictive capacity of our model.

\begin{table*}[ht]
\refstepcounter{table}\label{table measured}
\centering
{\fontsize{9}{9}\usefont{OT1}{phv}{m}{n}
\resizebox{\textwidth}{!}{\begin{tabular}{lcccccccccccccc}
\rowcolor{Gray}
\multicolumn{14}{l}{{\fontsize{11}{11}\usefont{OT1}{phv}{b}{n} Table~\thetable\ $\mid$} {\fontsize{10}{10}\usefont{OT1}{phv}{b}{n} MT dynamics parameters measured in \textit{Xenopus} egg extract spindles.}} \TopT\B \\
\rowcolor{LightGray}
\rule{-1pt}{3ex} & \textbf{Sawin} & \textbf{Waterman} & \textbf{Maddox} & \textbf{Tirnauer} & \textbf{Vallotton} & \textbf{Gaetz} & \textbf{Miyamoto} & \textbf{Yang} & \textbf{Needleman} & \textbf{Brugu\'{e}s} & \textbf{Reber} & \textbf{Decker} & \textbf{Dalton} \T \\
\rowcolor{LightGray}
 & \cite{sawin1991poleward} & \cite{waterman1998fluorescent} & \cite{maddox2003direct} & \cite{tirnauer2004microtubule} & \cite{vallotton2003recovery} & \cite{gaetz2004dynein} & \cite{miyamoto2004kinesin} & \cite{yang2008regional} & \cite{needleman2010fast} & \cite{brugues2012nucleation} & \cite{reber2013xmap215} & \cite{decker2018autocatalytic} & \cite{dalton2022gelation} \Tsmall\B \\
\hline
\rowcolor{LightGray}
$\tau_\text{MT}$ (s) & 75--100$^\dagger$ & --- & --- & --- & --- & --- & --- & --- & $16.0\pm1.5$ & --- & $18.2\pm2.1$ & $19.8\pm2.2^\ddagger$ & ${\sim}14.5$ \T \\
\rowcolor{LightGray}
$V_\parallel$ (\um/min) & $2.9\pm0.4$ & ${\sim}2$ & $2.3\pm0.6$ & ${\sim}2^\S$ & $2.5\pm0.2$ & $2.1\pm0.3$ & $2.21\pm0.45$ & $2.56\pm0.61$ & $2.17\pm0.62$ & $2.3\pm0.4$ & ${\sim}2^\S$ & --- & $2.5\pm0.4$ \Tsmall \\
\rowcolor{LightGray}
$V_\text{g}-V_\parallel$ (\um/min)$^\P$ & --- & --- & --- & $10.9\pm3.1$ & --- & --- & --- & --- & --- & ${\sim}10$ & $12.8\pm0.9$ & $23\pm8$ & --- \Tsmall\B \\
\hline
\rowcolor{LightGray}
\multicolumn{14}{l}{{\small $^\dagger$ Reported as a half-life $t_{1/2}$, corresponding to $\tau_\text{MT}=t_{1/2}/\ln 2\approx110$--$145$\,s; an outlier among the measurements that we do not consider.}} \T \\
\rowcolor{LightGray}
\multicolumn{14}{l}{{\small $^\ddagger$ Measured in monopolar spindles (kinesin-5 inhibited).\quad $^\S$ Value inferred from figures; not stated explicitly.}} \\
\rowcolor{LightGray}
\multicolumn{14}{l}{{\small $^\P$ Reported plus-end velocities in the laboratory frame; we add the corresponding sliding speed $V_\parallel$ to estimate the polymerization speed $V_\text{g}$.}} \B \\
\end{tabular}}}
\end{table*}

\textbf{Catastrophe rate}. The typical MT lifetime ($\tau_\text{MT} = 1/\lambda$) has been estimated primarily through speckle microscopy. In this technique, a small fraction of the tubulin pool is fluorescently labeled, providing a quantitative view of the dynamics of individual filaments within the dense MT network. Reported values for $\tau_\text{MT}$ in \emph{Xenopus} egg extract spindles are mostly in the range $15$--$20$\,s (Table~\ref{table measured}); an earlier study reported longer lifetimes~\cite{sawin1991poleward}.
These measurements underscore the striking contrast between the rapid turnover of individual MTs and the remarkable hours-long stability of the metaphase spindle, itself a scaffold built from these dynamic filaments.

\textbf{Sliding speed}. Direct measurements of crosslink-driven MT sliding velocities in \textit{Xenopus} egg extract spindles were also obtained through speckle microscopy. Reported values for $V_\parallel$ have been fairly consistent, in the range $2$--$3$\,\um/min (Table~\ref{table measured}).
Given the rapid catastrophe rate, these values of $V_\parallel$ indicate that a typical MT minus end translates a distance $V_\parallel / \lambda \sim 1$\,\um{} over its lifetime. This distance is a small fraction of the spindle's total length (about $30$--$50$\,\um), implying that MT minus ends nucleated near chromosomes, at the center of the spindle, would almost never reach the poles. This implies that the nucleation zone surrounding the chromosomes is broad \cite{burbank2007slide}.

\textbf{Polymerization speed}. In \textit{Xenopus} egg extract spindles, the velocity of polymerizing MT plus ends, essentially reflecting $V_\text{g}-V_\parallel$, has been measured by tracking comets of the plus-end tracking protein EB1. Reported values lie in the range $10$--$30$\,\um/min (Table~\ref{table measured}).
The larger variance in the reported measurements of $V_\text{g}$ is likely attributable to differences in experimental conditions. For instance, it has been shown that the viscosity of the cytoplasm can strongly affect the polymerization speed in yeast \cite{molines2022physical}. In mitotic astral spindles, a significant spatial variability in EB1-based measurements has also been observed \cite{redemann2017c}. Taking a moderate estimate of $V_\text{g} \sim 12.5$\,\um/min and $\tau_\text{MT} \sim 20$\,s, it can be deduced that a typical MT reaches a length of $\ell=V_\text{g} / \lambda \sim 4$\,\um, roughly consistent with laser ablation experiments \cite{brugues2012nucleation}.

\section{Numerical scheme}
\label{app:numerical-scheme}

We solve the reduced-dimension dynamical system using a pseudo-spectral method on a 1D periodic domain of length $L$, discretized with $N$ uniformly spaced points. Spatial derivatives are evaluated spectrally, with the Nyquist mode zeroed throughout and a 2/3 dealiasing rule applied to nonlinear evaluations.

\subsection{Nondimensionalization}
\label{app:nondim}
Time is measured in units of $1/\lambda$ and space in units of $\ell=V_\text{g}/\lambda$, so that the dimensionless growth speed is $1$ and the sliding speed is $\varepsilon = V_\parallel/V_\text{g}$. Dimensionless parameters are marked with a tilde: $\tilde{D} = D/(\ell^2\lambda)$, $\tilde{\lambda}_c = \lambda_c/\lambda$, $\tilde{n}_b = n_b/\lambda$, $\tilde{\ell}_s = \ell_s/\ell$, and $\tilde{a}_c = a_c/\lambda$; the ratios $a_0$, $a_m$, $\varepsilon$, $s_0$, and the integer $M$ are already dimensionless.
In these units, the equations for the right- and left-pointing populations become
\begin{align}
    \partial_t\eta_\pm \pm (1-\varepsilon)\partial_x\eta_\pm &= n_\pm - \eta_\pm, \label{eq num eta}\\
    \partial_t\phi^{(0)}_\pm \mp \varepsilon\partial_x\phi^{(0)}_\pm &= a_0 \eta_\pm - \phi^{(0)}_\pm, \label{eq num phi0}\\
    \partial_t\phi^{(m)}_\pm \mp \varepsilon\partial_x\phi^{(m)}_\pm &= a_m\phi^{(m-1)}_\pm - \phi^{(m)}_\pm, \quad m=1,\dots,M, \label{eq num phim}
\end{align}
where $a_0=\ell/ \ell_0$, $a_m=k_\text{on}^{(m)}R^{(m)}/\lambda$ (for $m\geq1$) and $n_\pm=\tilde{n}_b\, c_\pm$. The total nucleator concentration satisfies
\begin{equation}
    \partial_t c + \varepsilon\partial_x(c_{-} - c_{+}) - \tilde{D}\partial_x^2 c_\text{u} = -\tilde{\lambda}_c c + f_\text{act}(s), \label{eq num c}
\end{equation}
with
\[c_\text{u} = c/\bigl(1+\alpha(\phi^{(M)}_+ + \phi^{(M)}_-)\bigr), \qquad c_\pm = \alpha\phi^{(M)}_\pm c_\text{u}.
\]

\subsection{Time integration}
\label{app:time-integration}
Each equation is written in the form $\partial_t \hat{u} = \mathcal{L}\hat{u} + \mathcal{N}$, where $\mathcal{L}\hat{u}$ collects same-field linear terms treated via an integrating factor and $\mathcal{N}$ collects the terms treated explicitly: nonstiff/nonlinear terms and sources defined by preceding fields in the chain. In Fourier space with wavenumber $k$,
\begin{alignat}{3}
  &\text{for }\eta_\pm\text{:}\qquad              & \mathcal{L} &= \mp(1-\varepsilon)ik - 1, \qquad & \mathcal{N} &= \hat{n}_\pm, \\
  &\text{for }\phi^{(0)}_\pm\text{:}\qquad           & \mathcal{L} &= \pm\varepsilon ik - 1,    \qquad & \mathcal{N} &= a_0\,\hat{\eta}_\pm, \\
  &\text{for }\phi^{(m)}_\pm,\,m\!\geq\!1\text{:}\qquad & \mathcal{L} &= \pm\varepsilon ik - 1, \qquad & \mathcal{N} &= a_m\,\hat{\phi}_\pm^{(m-1)}, \\
  &\text{for }c\text{:}\qquad                        & \mathcal{L} &= -\tilde{D} k^2 - \tilde{\lambda}_c,    \qquad & \mathcal{N} &= \varepsilon ik(\hat{c}_+-\hat{c}_-) - \tilde{D} k^2(\hat{c}_\text{u}-\hat{c}) + \hat{f}_\text{act}(s).
\end{alignat}
For the $\eta$, $\phi^{(0)}$, and $\phi^{(m)}$ fields, both advection and decay are absorbed into $\mathcal{L}$. For $c$, the diffusion term in the governing equation acts on $c_\text{u}$ (a nonlinear function of $c$ and $\phi_\pm^{(M)}$) rather than on $c$ itself. We therefore split $\tilde{D}\partial_x^2 c_\text{u} = \tilde{D}\partial_x^2 c + \tilde{D}\partial_x^2(c_\text{u}-c)$: the linear part $\tilde{D}\partial_x^2 c$ is absorbed into $\mathcal{L}$ to remove diffusive stiffness, while the correction $\tilde{D}\partial_x^2(c_\text{u}-c)$ and the advection term are treated explicitly in $\mathcal{N}$.

For each field, we use the precomputed integrating factor $\mathrm{IF} = e^{\mathcal{L}\Delta t}$ and apply the IFAB2 update (second-order Integrating Factor Adams--Bashforth method) \cite{cox2002exponential}:
\begin{equation}
    \hat{u}^{n+1} = \text{IF}\,\hat{u}^n + \text{IF}\,\frac{\Delta t}{2}\left(3\mathcal{N}^n - \text{IF}\mathcal{N}^{n-1}\right). \label{eq IFAB2}
\end{equation}
The scheme is started with a forward Euler step. Initial conditions are $c=0$ and a small uniform $\eta_\pm$; the fields $\phi^{(0)}_\pm,\dots,\phi^{(M)}_\pm$ are initialized by setting $\partial_t=0$ in each equation and solving the resulting chain recursively.

\subsection{Parameters}
\label{app:parameters}

All simulations use the dimensionless model (Appendix~\ref{app:nondim}), with time in units of $1/\lambda$ and length in units of $\ell = V_\text{g}/\lambda$. The chosen parameter values are rationalized below and summarized in Table~\ref{table params}.

\textit{Intrinsic MT parameters.} The growth speed and catastrophe rate are fixed at $1$ by nondimensionalization. The sliding ratio $\varepsilon = V_\parallel/V_\text{g}$ is one of the two control parameters of this study. Its experimental reference value $\varepsilon \approx 0.2$ follows from the measured velocities $V_\parallel \approx 2.5$\,\um/min and $V_\text{g} \approx 12.5$\,\um/min (Table~\ref{table measured}).

\textit{Branching complex assembly and nucleation.} We fixed the nucleation prefactor to $\tilde{n}_b = 270$ in order to sustain approximately 1000 simultaneous filaments at steady state. The lattice-site production rate $a_0 = \ell/\ell_0 = 1000$ is fixed by molecular geometry (mean MT length $\ell \approx 4$--$5$\,\um, tubulin monomer size $\ell_0 \approx 4$\,nm~\cite{mandelkow1986surface}). The multistep association rate is unknown; we fixed it to $a_m =k_\text{on}^{(m)}R^{(m)}/\lambda =0.1$ to be consistent with the rate-limiting condition $k_\text{on}^{(m)}R^{(m)} \ll \lambda$ (Appendix~\ref{app:rate-limiting}). The number of assembly steps $M$ is the second key control parameter and is varied per figure.

\textit{Nucleator species ($c$).} The deactivation rate $\tilde{\lambda}_c = 0.2$ gives the nucleator a lifetime five times the MT turnover time. The diffusion unit is $\ell^2\lambda \approx 0.8$\,\um$^2$/s, so setting $\tilde{D}=7.5$ corresponds to a physical diffusivity $\tilde{D}\,\ell^2\lambda \approx 6$\,\um$^2$/s, within the range $3$--$8$\,\um$^2$/s measured by FRAP for $\gamma$-TuRC~\cite{endow2011anastral}. The binding affinity $\alpha$ of the nucleator to the mature lattice complex is unknown; 
we treat it as a fitting parameter, rescaled to hold the amplitude loop gain fixed for different values of $n_b$, $a_0$, $a_m$, and $M$. Specifically, we use $\alpha=20/(\tilde{n}_b a_0 a_m^M)$ in the dimensionless simulations.

\textit{RanGTP gradient and activation.} The chromosomal signal $s(x) = e^{-|x|/\tilde{\ell}_s}$ with decay length $\tilde{\ell}_s = 0.5$ approximates a localized RanGTP gradient, and drives the activation of $c$ through $f_\text{act}(s) = \tilde{a}_c\, s/(s_0+s)$ with half-saturation $s_0 = 0.01$. Choosing $s_0 \ll s_{\max}=1$ ensures that activation is near-saturating (switch-like) across a broad midzone, while the $\tilde{a}_c=\tilde{\lambda}_c$ prefactor keeps the peak steady-state nucleator concentration $O(1)$. Keeping $c(x)$ flat across a wide midzone is important for controlling the spindle size.

\textit{Space/time discretization.} We use domain size $L = 40$, $N = 1024$ grid points, timestep $\Delta t = 0.02$, and total simulation time $T_\text{final} = 200$, all in dimensionless units. The domain is taken much larger than the spindle size so that MT densities decay to negligible values before reaching the periodic boundaries; all figures display only the central segment of the computational domain, to highlight the condensate.

\begin{table*}[ht]
\refstepcounter{table}\label{table params}
\centering
{\fontsize{9}{9}\usefont{OT1}{phv}{m}{n}
\resizebox{\textwidth}{!}{\begin{tabular}{llll}
\rowcolor{Gray}
\multicolumn{4}{l}{{\fontsize{11}{11}\usefont{OT1}{phv}{b}{n} Table~\thetable\ $\mid$} {\fontsize{10}{10}\usefont{OT1}{phv}{b}{n} Simulation parameters.}} \TopT\B \\
\rowcolor{LightGray}
\textbf{Parameter} & \textbf{Description} & \textbf{Value} & \textbf{Remarks} \T\B \\
\hline
\rowcolor{LightGray}
\multicolumn{4}{l}{\textit{Intrinsic MT parameters}} \T\B \\
\hline
\rowcolor{LightGray}
$\lambda$ & Catastrophe rate & $1$ & Fixed by nondimensionalization \T \\
\rowcolor{LightGray}
$V_\text{g}$ & Growth speed & $1$ & Fixed by nondimensionalization \Tsmall \\
\rowcolor{LightGray}
$\varepsilon = V_\parallel/V_\text{g}$ & Sliding ratio & varied & Control parameter; reference $0.2$ ($V_\parallel \approx 2.5$, $V_\text{g} \approx 12.5$\,\um/min) \Tsmall\B \\
\hline
\rowcolor{LightGray}
\multicolumn{4}{l}{\textit{Branching complex assembly and nucleation parameters}} \T\B \\
\hline
\rowcolor{LightGray}
$\tilde{n}_b=n_b/\lambda$ & Nucleation prefactor & $270$ & Free; sets total filament count \T \\
\rowcolor{LightGray}
$a_0 = \ell/\ell_0$ & Lattice-site production rate & $1000$ & Molecular geometry ($\ell \approx 4\text{--}5$\,\um, $\ell_0 \approx 4$\,nm) \Tsmall \\
\rowcolor{LightGray}
$a_m = k_\text{on}^{(m)}R^{(m)}/\lambda$ & Multistep association rate & $0.1$ & Rate-limiting condition $a_m \ll 1$ (Appendix~\ref{app:rate-limiting}) \Tsmall \\
\rowcolor{LightGray}
$M$ & Number of assembly steps & varied & Control parameter \Tsmall\B \\
\hline
\rowcolor{LightGray}
\multicolumn{4}{l}{\textit{Nucleator species ($c$) parameters}} \T\B \\
\hline
\rowcolor{LightGray}
$\tilde{D}=D/(\ell^2\lambda)$ & Nucleator diffusivity & $7.5$ & $\gamma$-TuRC FRAP: ${\sim}6$\,\um$^2$/s \cite{endow2011anastral} \T \\
\rowcolor{LightGray}
$\tilde{\lambda}_c=\lambda_c/\lambda$ & Nucleator deactivation rate & $0.2$ & Free; lifetime $\sim\!5\times$ MT turnover time \Tsmall \\
\rowcolor{LightGray}
$\alpha$ & Nucleator binding affinity & $20/(\tilde{n}_b a_0 a_m^M)$ & Rescaled $\propto 1/(\tilde{n}_b\, a_0\, a_m^{M})$ so $\tilde{n}_b$ alone sets filament count \Tsmall\B \\
\hline
\rowcolor{LightGray}
\multicolumn{4}{l}{\textit{RanGTP gradient and activation parameters}} \T\B \\
\hline
\rowcolor{LightGray}
$\tilde{\ell}_s=\ell_s/\ell$ & RanGTP gradient decay length & $0.5$ & Free; localized source ($\tilde{\ell}_s \ll L$) \T \\
\rowcolor{LightGray}
$\tilde{a}_c=a_c/\lambda$ & Activation plateau & $0.2$ & Keeps peak nucleator level $c = O(1)$ \Tsmall \\
\rowcolor{LightGray}
$s_0$ & Activation half-saturation & $0.01$ & $s_0 \ll s_{\max}$ (switch-like) \Tsmall\B \\
\hline
\rowcolor{LightGray}
\multicolumn{4}{l}{\textit{Space/time discretization}} \T\B \\
\hline
\rowcolor{LightGray}
$L$ & Domain size & $40$ &  $>$ spindle size \T \\
\rowcolor{LightGray}
$N$ & Number of grid points & $1024$ & $\Delta x=L/N\ll 1$ MT length \Tsmall \\
\rowcolor{LightGray}
$\Delta t$ & Timestep & $0.02$ & $\ll 1/\lambda$; numerically stable \Tsmall \\
\rowcolor{LightGray}
$T_\text{final}$ & Simulation duration & $200$ & $\gg 1/\lambda$; long enough to integrate transient behavior \Tsmall\B \\
\hline
\end{tabular}}}
\end{table*}

\section{Stochastic filament simulations}
\label{app:gillespie}

To provide a filament-level view of the continuum model solutions, we simulate individual MTs as a birth--death process driven by the numerically evolved nucleation fields $n_\pm(x,t)$ of the continuum problem. These Gillespie-type simulations do not involve explicit branching of one filament from another, nor any interfilament interactions; their purpose is visualization. Resulting time-lapse dynamics are provided in Figs.~\ref{fig:dynamics} and~\ref{fig:dynamics-M5} and in Movies~\ref{mov:M0-eps02}--\ref{mov:M1-eps02}. All quantities are in dimensionless units. 

\textit{Time-dependent births.} To visualize the transient dynamics, we utilize snapshots of $n_\pm(x,t)$ saved by the PDE solver at uniform intervals $\Delta t_\text{save}$, producing the sequence $n_\pm^{(k)}(x) \equiv n_\pm(x,\,k\,\Delta t_\text{save})$. Within frame $k$, covering $[t_k,\,t_k+\Delta t_\text{save})$, births are sampled from $n_\pm^{(k)}(x)$ held fixed. Let $R_\pm^{(k)} = \int n_\pm^{(k)}(x)\,dx$ be the total instantaneous nucleation rate for each polarity, and $R_\text{nuc}^{(k)} = R_+^{(k)} + R_-^{(k)}$ the combined rate. Inter-arrival times $\delta t \sim \mathrm{Exp}(R_\text{nuc}^{(k)})$ are drawn until the frame interval is exhausted, and each event is assigned a polarity ($\pm$) with probability $R_\pm^{(k)}/R_\text{nuc}^{(k)}$ (typically equaling $1/2$) and a position $x_b$ drawn from the respective PDF $n_\pm^{(k)}(x)/R_\pm^{(k)}$. This piecewise-constant scheme converges to the time-dependent Poisson process as $\Delta t_\text{save}\to 0$.

\textit{Deterministic horizontal kinematics.} A filament born with polarity $\pm$ and position $x_b$ at time $t_b$ evolves deterministically: its minus end drifts as $x_b \mp \varepsilon(t-t_b)$ and its length grows as $l = t - t_b$. Accordingly, its plus end advances as $x_b \pm (1-\varepsilon)(t-t_b)$.

\textit{Probabilistic y-positions.} Each newborn filament is assigned a transverse coordinate $y_b$ drawn uniformly in a fixed display interval and held constant throughout.

\textit{Probabilistic deaths.}
Upon birth, each filament is assigned an independent lifetime $\tau \sim \mathrm{Exp}(1)$ (recalling $\lambda=1$ in dimensionless units), after which it is removed. Filaments whose lifetime has elapsed are pruned at the start of each frame.

\section{The rate-limiting assumption}
\label{app:rate-limiting}
The polynomial distribution of complexes (growing from plus end to minus end), given by the kernels of Eqs.~\eqref{eq phi m plus} and~\eqref{eq phi m minus}, rests on two key assumptions:
(i) Branching complexes, at any stage of assembly, do not disassemble within the MT lifetime.
(ii) Each assembly step is slow enough that intermediates remain far from saturation. In other words, complexes accumulate gradually, limited by conversion kinetics rather than binding-site availability.

We formalize these conditions in the reduced-dimension framework, starting from a generalized assembly model that includes backward transitions at rates $k_{\text{off}}^{(m)}$, depletion of stage $m$ upon conversion to $m+1$, and consumption of mature complexes upon nucleation.

\noindent For $m = M$ (final state), we have
\begin{equation}
\partial_t\phi^{(M)}_+(x,t)-V_\parallel\partial_x\phi^{(M)}_+
= k_{\text{on}}^{(M)}R^{(M)}\phi^{(M-1)}_+ 
- \big(k_{\text{off}}^{(M)} + \lambda\big)\phi^{(M)}_+-n_+[\phi_+^{(M)}]. \label{eq M reversible}
\end{equation}
The consumption term $-n_+$ applies when each nucleation event uses up a mature complex. If, however, mature complexes act as reusable catalytic platforms, this term is omitted: after nucleating a daughter MT, a platform remains available for further nucleation on the mother lattice. 

\noindent For $1 \leq m \leq M-1$ (intermediate states):
\begin{align}
\partial_t\phi^{(m)}_+(x,t)-V_\parallel\partial_x\phi^{(m)}_+
=\, & k_{\text{on}}^{(m)}R^{(m)}\phi^{(m-1)}_+ 
- \big(k_{\text{off}}^{(m)} + \lambda\big)\phi^{(m)}_+ \nonumber\\
&  - k_{\text{on}}^{(m+1)}R^{(m+1)}\phi_+^{(m)}
+ k_{\text{off}}^{(m+1)}\phi_+^{(m+1)}. \label{eq intermediate m reversible}
\end{align}

\noindent For $m = 0$ (bare polymerized tubulin):
\begin{equation}
\partial_t\phi^{(0)}_+(x,t)-V_\parallel\partial_x\phi^{(0)}_+
= \frac{V_\text{g}}{\ell_0}\eta_+ - \lambda\phi^{(0)}_+
- k_{\text{on}}^{(1)}R^{(1)}\phi^{(0)}_+ 
+ k_{\text{off}}^{(1)}\phi_+^{(1)}. \label{eq m=0 reversible}
\end{equation}

\noindent For plus ends ($\eta_+$), the formulation is unchanged:
\begin{equation}
    \partial_t\eta_+(x,t)+(V_\text{g}-V_\parallel)\partial_x\eta_+= n_+[\phi_+^{(M)}]
    -\lambda\eta_+. \label{eq eta reversible}
\end{equation}

We now identify the limit of Eqs.~\eqref{eq M reversible}--\eqref{eq eta reversible} that recovers the irreversible forward model, Eqs.~\eqref{eq phi plus m},~\eqref{eq phi pm 0}, and~\eqref{eq eta pm}, consistent with the polynomial Ansatz Eqs.~\eqref{eq phi m plus} and~\eqref{eq phi m minus}. Specifically, we seek the parameter regime under which the RHS of Eq.~\eqref{eq intermediate m reversible} reduces to
\[
k_{\text{on}}^{(m)}R^{(m)}\phi^{(m-1)}_+ 
- \lambda\phi^{(m)}_+.
\]
Backward transitions are negligible when:
\begin{equation}
k_\text{off}^{(m)}\phi_+^{(m)}\ll k_\text{on}^{(m)}R^{(m)}\phi_{+}^{(m-1)} \quad \&\quad k_\text{off}^{(m)}\ll \lambda.
\end{equation}
Using the local quasi-steady-state estimate $\phi^{(m)}\simeq (k_\text{on}^{(m)}R^{(m)}/\lambda)\phi^{(m-1)}$, applicable when precursor fields vary slowly along lattice trajectories over the turnover time, both conditions are satisfied when $k_\text{off}^{(m)}\ll\lambda$.
Forward conversion loss is negligible when:
\begin{equation}
k_\text{on}^{(m+1)}R^{(m+1)}\phi_+^{(m)}\ll k_\text{on}^{(m)}R^{(m)}\phi_+^{(m-1)}\quad \& \quad k_\text{on}^{(m+1)}R^{(m+1)} \ll \lambda.
\end{equation}
The same estimate gives $k_\text{on}^{(m+1)}R^{(m+1)} \ll \lambda$ as the condition for negligible forward conversion loss.

In summary, assuming reusable catalytic platforms, we recover the forward buildup model to leading order under the kinetic assumptions:
\begin{equation}
k_\text{off}^{(m)}\ll \lambda \quad \& \quad k_\text{on}^{(m)}R^{(m)}\ll\lambda \quad \text{for }m\geq1.
\end{equation}
These conditions make each assembly step slow relative to MT turnover and therefore rate-limiting. Over a typical MT lifetime, each intermediate undergoes little depletion through onward conversion or disassembly. Forward recruitment nevertheless supplies the leading nonzero contribution to each successive assembly stage, while depletion and reverse transitions contribute higher-order corrections. Assembly thus proceeds through gradual accumulation rather than equilibration: each stage integrates the buildup of the preceding stage over the residence time of a lattice site. This successive integration gives the polynomial residence-time kernels of the forward buildup model to leading order. Older lattice sites, located toward the mother's minus end, consequently host more mature complexes, producing the minus-end bias in Eqs.~\eqref{eq phi m plus} and~\eqref{eq phi m minus}.

\section{Branchpoint drift in a background gradient of nucleators}
\label{app:gradient-drift}

Here, we extend the analysis of Section~\ref{sec:drift} to explicitly account for spatial variations in the concentration of the Ran-activated species, $c(x)$. Recall that we treat the binding of this species to mature ($M$-staged) complexes as the final step in the branching reaction. Accordingly, we take the nucleation rate along the parent filament as $n_+(s) \propto (L - s)^M c(x)$, with $x = X + s$. This form follows Eq.~\eqref{eq n m step} more closely, while still neglecting spatial variations in the denominator. Relative to the current minus end $X$, the center of mass of this expression is given by
\[
X^{(n)}-X= \frac{\int_0^L s(L - s)^Mc(X+s)\, ds}{\int_0^L (L - s)^Mc(X+s)\, ds}.
\]
We now assume $c$ is slowly varying and expand it around $x_0$ (the nucleation point of the parent MT):
\[c(X+s)=c(x_0-V_\parallel\tau+s)\approx c(x_0)+(-V_\parallel\tau+s)c'(x_0).\]
Then, the numerator gives
\[
\int_0^L s(L - s)^Mc(X+s)\, ds\approx\frac{L^{M+2}}{(M+1)(M+2)}\left(c(x_0)-V_\parallel\tau\,c'(x_0)\right) + \frac{2 L^{M+3}}{(M+1)(M+2)(M+3)}c'(x_0).
\]
The denominator gives
\[
\int_0^L (L - s)^Mc(X+s)\, ds\approx\frac{L^{M+1}}{M+1}\left(c(x_0)-V_\parallel\tau\,c'(x_0)\right) + \frac{ L^{M+2}}{(M+1)(M+2)}c'(x_0).
\]
This yields
\[
X^{(n)}-X\approx \frac{L}{M+2}+\frac{c'(x_0)}{c(x_0)}\frac{(M+1)L^2}{(M+2)^2(M+3)}.
\]
Substituting $X(\tau)$ and $L(\tau)$:
\begin{equation}
    X^{(n)}(\tau)\approx x_0+\left(-V_\parallel+\frac{V_\text{g}}{M+2}\right)\tau+\frac{c'(x_0)}{c(x_0)}\frac{(M+1)V_\text{g}^2}{(M+2)^2(M+3)}\tau^2. \label{eq:Xn gradient}
\end{equation}

Now, let us consider that our parent right-pointing filament was nucleated to the left of the origin in the laboratory frame, i.e., $x_0<0$ (as expected for a spindle at steady state). Because the saturating activation $f_\text{act}$ renders $c(x)$ nearly flat across the interior of the RanGTP zone, $c'(x_0)\approx0$ there and the gradient term is negligible; it becomes appreciable only near the zone edges, where $c$ falls off (with $c'(x_0)>0$ for $x_0<0$). Where it is appreciable, a negative drift of successive nucleation events requires the backward sliding term $-V_\parallel$ to overcome not only the forward velocity contribution $V_\text{g}/(M+2)$ but also the positive quadratic term, factored by $c'(x_0)/c(x_0)$.

A backward mean displacement would correspond to $\langle X^{(n)}\rangle < x_0$, where $\langle\cdot\rangle$ denotes an average over the parent-age distribution of branching events. Because $X^{(n)}(\tau)$ contains a quadratic term, simply evaluating it at the mean parent age, i.e., $X^{(n)}(\tau_n)$, does not give the mean position. Furthermore, the nucleator gradient changes the relative contribution of different parent ages to branching. This means that the Gamma distribution for $\tau$, obtained in Section~\ref{sec:mean-branching-time} under the assumption of a uniform background nucleator, must also be corrected. To first order in the nucleator gradient, the corrected distribution $g(\tau)$ is proportional to
\begin{align}
g(\tau)\propto&\, e^{-\lambda \tau}\int_0^L (L-s)^M c(X+s)\, ds \Big|_{L=V_\text{g}\tau} \nonumber
\\
\simeq&\, \frac{V_\text{g}^{M+1}}{M+1}\left(c(x_0)-V_\parallel\tau c'(x_0)\right)\tau^{M+1} e^{-\lambda \tau}+\frac{V_\text{g}^{M+2}\,c'(x_0)}{(M+1)(M+2)}\tau^{M+2} e^{-\lambda \tau}\nonumber\\
=&\, \frac{V_\text{g}^{M+1} c(x_0)}{M+1}\left[\tau^{M+1}e^{-\lambda \tau} +\frac{c'(x_0)}{c(x_0)}\left(-V_\parallel+\frac{V_\text{g}}{M+2}\right) \tau^{M+2}e^{-\lambda \tau} \right].\nonumber
\end{align}
The first term is proportional to the uncorrected Gamma distribution of shape parameter $M+2$, while the second term is a small correction proportional to the Gamma distribution of one higher order (shape parameter $M+3$). A positive correction term therefore shifts the distribution toward longer parent ages, whereas a negative correction shifts it toward shorter ages. Interestingly, the sign of this term is determined by $c'(x_0)\left(-V_\parallel + V_\text{g}/(M+2)\right)$, which is $c'(x_0)$ times the intergenerational drift velocity to leading order. This agrees with intuition: a drift of likely branching sites toward higher nucleator concentrations favors branching at older parent ages, whereas their drift toward lower concentrations favors branching at younger ages. Consistently, the mean parent age sampled across branching events is
\[
\langle\tau\rangle_g = \frac{\int_0^\infty \tau g(\tau)\, d\tau}{\int_0^\infty g(\tau)\, d\tau} \simeq \frac{M+2}{\lambda} + \frac{M+2}{\lambda^2}\frac{c'(x_0)}{c(x_0)} \left(-V_\parallel+\frac{V_\text{g}}{M+2}\right).
\]

To evaluate $\langle X^{(n)}\rangle_g$, we also need the second
moment $\langle\tau^2\rangle_g$ of the parent age. In Eq.~\eqref{eq:Xn gradient}, however,
that quadratic term is already multiplied by the small factor
$c'(x_0)/c(x_0)$. It therefore suffices to use the second moment
of the leading-order Gamma distribution $f(\tau;\,M{+}2,\lambda)$,
\[
\langle\tau^2\rangle_f= \frac{(M+2)(M+3)}{\lambda^2}.
\] 
Putting everything together, we find, to first order in the nucleator gradient,
\begin{align}
\langle X^{(n)}\rangle_g &\simeq x_0 + \left(-V_\parallel + \frac{V_\text{g}}{M+2}\right)\langle\tau\rangle_g + \frac{c'(x_0)}{c(x_0)}\frac{(M+1)V_\text{g}^2}{(M+2)^2(M+3)}\langle\tau^2\rangle_f \nonumber\\
&= x_0 + \frac{V_\text{g}-(M+2)V_\parallel}{\lambda}
+ \frac{c'(x_0)}{c(x_0)}\frac{1}{\lambda^2}
\left[
(M+2)\left(-V_\parallel+\frac{V_\text{g}}{M+2}\right)^2
+\frac{M+1}{M+2}V_\text{g}^2
\right] \nonumber\\
&= x_0 + \frac{V_\text{g}-(M+2)V_\parallel}{\lambda}
+ \frac{c'(x_0)}{c(x_0)}
\frac{\left(V_\text{g}-V_\parallel\right)^2 + (M+1) V_\parallel^2}
{\lambda^2}.
\end{align}
In the last term, the numerator multiplying the gradient is strictly positive, so the correction carries the sign of $c'(x_0)$: branching is pulled toward higher nucleator concentration, as expected. To first order in the nucleator gradient, the condition for a negative intergenerational drift $\langle X^{(n)}\rangle<x_0$ becomes
\begin{equation}
\frac{V_\parallel}{V_\text{g}}
>
\frac{1}{M+2}
+\frac{M+1}{(M+2)^2}
\frac{c'(x_0)}{c(x_0)}
\frac{V_\text{g}}{\lambda}.
\label{eq:gradient-drift-condition}
\end{equation}
A positive gradient therefore raises the sliding-to-growth velocity ratio required for backward drift, as occurs to the left of the origin where the nucleator concentration rises toward the center. To the right of the origin, where $c'(x_0)<0$, the gradient instead enhances leftward drift of the right-pointing population.

\section{Cumulative parent-age distribution}
\label{app:branching-probability}
In a balanced steady-state population, each MT produces, on average, one daughter. The Gamma distribution $f(\tau;\,M{+}2,\lambda)$ introduced in Section~\ref{sec:mean-branching-time} gives the PDF of parent age $\tau$ at which a random branching event occurs. Its CDF,
\[
F(\tau;\,M{+}2,\lambda) := \int_0^\tau f(u;\,M{+}2,\lambda)\,du,
\]
gives the fraction of branching events occurring on parents younger than $\tau$. Since $M+2$ is an integer, this reduces to the Erlang form:
\begin{equation}
F(\tau;\,M{+}2,\lambda)
=1-e^{-\lambda\tau}\sum_{m=0}^{M+1}\frac{(\lambda\tau)^m}{m!}.
\label{eq:erlang-cdf}
\end{equation}
The fraction of branching events occurring before the parent reaches the mean MT lifetime, \(\lambda^{-1}\), is thus
\begin{equation}
r(M)=F(\lambda^{-1};\,M{+}2,\lambda)
=1-e^{-1}\sum_{m=0}^{M+1}\frac{1}{m!}.
\label{eq:rM}
\end{equation}
For $M=0$: $r(0)=1-2/e\approx0.264$; for $M=1$: $r(1)\approx0.08$; for $M=4$: $r(4)\approx6\times10^{-4}$. Hence at $M=4$, about $99.94\%$ of branching events occur on parents older than $1/\lambda$. 

Similarly, the fraction of branching events occurring on parents younger than $2/\lambda$, the mean parent age at branching in the single-step model, is
\begin{equation}
r_2(M)=F(2\lambda^{-1};\,M{+}2,\lambda)
=1-e^{-2}\sum_{m=0}^{M+1}\frac{2^m}{m!}.
\label{eq:r2M}
\end{equation}
For $M=0$: $r_2(0)\approx0.594$; for $M=1$: $r_2(1)\approx0.323$; for $M=4$: $r_2(4)\approx0.017$. The drop across just a few steps in $M$ is stark, reflecting how sharply the multistep mechanism focuses branching onto a small fraction of exceptionally long-lived filaments.

\section{Cooperative recruitment: a close relative to multistep assembly}
\label{app:cooperative}
Here, we discuss a closely related, and perhaps complementary, mechanism to multistep branching that also yields a strong minus-end bias in branching events, thereby enabling a negative intergenerational drift. To illustrate this mechanism, we consider only one slow intermediary of density $\phi^{(1)}_\pm$ that binds to $\phi_\pm^{(0)}$ and serves as a platform for the Ran-activated species $c$. The key idea is that $\phi^{(1)}_\pm$ also binds onto \emph{itself}. Formally:
\begin{equation}
    \partial_t\phi^{(1)}_+(x,t)-V_\parallel\partial_x\phi^{(1)}_+=k_\text{on}^{(1)}R^{(1)}\phi^{(0)}_++(k_\text{self}^{(1)}R^{(1)}-\lambda)\phi^{(1)}_+, \label{eq: dt phi1 self}
\end{equation}
where $k_\text{self}^{(1)}$ is the rate of cooperative binding. It is clear from this equation that the inequality $k_\text{self}^{(1)}R^{(1)}<\lambda$ must hold to avoid unbounded growth of $\phi^{(1)}_+$. 

Given the full MT number density, $\psi_+(x,l,t)$, we make the following Ansatz:
\begin{equation}
    \phi^{(1)}_+(x,t)=\frac{A_1}{\ell_0}\int_0^\infty dl \int_0^l ds \frac{e^{a_\text{s}\frac{l-s}{V_\text{g}}}-1}{a_\text{s}}\psi_+(x-s,l,t)=:\mathcal{H}_+^{(1)}[\psi_+], \label{eq phi1 self def}
\end{equation}
where $A_1=k_\text{on}^{(1)}R^{(1)}$, as before, and we defined $a_\text{s}=k_\text{self}^{(1)}R^{(1)}$. 
As a first sanity check, we note that in the limit $k_\text{self}^{(1)}\to0$ (Eq.~\eqref{eq: dt phi1 self} becomes Eq.~\eqref{eq phi plus m}) one has:
 \[
\frac{e^{a_\text{s}\frac{l-s}{V_\text{g}}}-1}{a_\text{s}}\approx \frac{l-s}{V_\text{g}}. 
 \]
 Then, Eq.~\eqref{eq phi1 self def} agrees with Eq.~\eqref{eq phi m plus} for $m=1$, i.e., $\mathcal{H}_+^{(1)}\to\mathcal{L}_+^{(1)}$, as expected.
 
To verify that Eq.~\eqref{eq phi1 self def} indeed satisfies Eq.~\eqref{eq: dt phi1 self}, we apply $\mathcal{H}_+^{(1)}$ to Eq.~\eqref{eq kinetic growth} term by term:
\begin{align}
    &\mathcal{H}_+^{(1)}[\partial_t\psi_+]=\partial_t\mathcal{H}_+^{(1)}[\psi_+]=\partial_t\phi^{(1)}_+, \\
    &\mathcal{H}_+^{(1)}[-V_\parallel\partial_x\psi_+]=-V_\parallel\partial_x\mathcal{H}_+^{(1)}[\psi_+]=-V_\parallel\partial_x \phi^{(1)}_+, \\
    &\mathcal{H}_+^{(1)}[V_\text{g}\partial_l\psi_+]= \frac{A_1}{\ell_0} \int_0^\infty dl\int_0^lds\frac{e^{a_\text{s}\frac{l-s}{V_\text{g}}}-1}{a_\text{s}}V_\text{g}\partial_l\psi_+(x-s,l) \nonumber \\
    & \qquad\quad =\frac{A_1}{\ell_0}\int_0^\infty ds\int_{s}^\infty dl \frac{e^{a_\text{s}\frac{l-s}{V_\text{g}}}-1}{a_\text{s}}V_\text{g}\partial_l\psi_+(x-s,l) \nonumber \\
    & \qquad\quad =\frac{A_1}{\ell_0}\int_0^\infty ds\Bigg( \left[\frac{e^{a_\text{s}\frac{l-s}{V_\text{g}}}-1}{a_\text{s}}V_\text{g}\psi_+(x-s,l)\right]^{l=\infty}_{l=s}  
    -\int_{s}^\infty dl \,e^{a_\text{s}\frac{l-s}{V_\text{g}}} \psi_+(x-s,l)\Bigg) \nonumber \\
    & \qquad\quad  
    =-\frac{A_1}{\ell_0}\int_0^\infty dl\int_{0}^l ds \left(a_\text{s}\frac{e^{a_\text{s}\frac{l-s}{V_\text{g}}}-1}{a_\text{s}}+1\right) \psi_+(x-s,l) \nonumber \\
    & \qquad \quad = -a_\text{s} \mathcal{H}_+^{(1)}[\psi_+]-k_\text{on}^{(1)}R^{(1)}\mathcal{L}_+^{(0)}[\psi_+] = - k_\text{self}^{(1)}R^{(1)}\phi_+^{(1)} -k_\text{on}^{(1)}R^{(1)}\phi^{(0)}_+, \nonumber
\end{align}
where we considered $\psi_+\sim e^{-\lambda l/V_\text{g}}$ and assumed $k_\text{self}^{(1)}R^{(1)}<\lambda$. Thus, assembling all terms together yields exactly Eq.~\eqref{eq: dt phi1 self}.

\subsection{Relationship to the multistep model}
\label{app:cooperative-multistep}
Equation~\eqref{eq phi1 self def} of the cooperative model, like Eq.~\eqref{eq phi m plus} of the multistep model, confers an arbitrarily strong minus-end bias of the branching complex. In fact, these models are closely related algebraically; Taylor expanding the kernel in Eq.~\eqref{eq phi1 self def} yields:
\[
\frac{A_1}{\ell_0}\frac{e^{a_\text{s}\frac{l-s}{V_\text{g}}}-1}{a_\text{s}}=\frac{k_\text{on}^{(1)}R^{(1)}}{\ell_0}\sum_{m=1}^{\infty}\frac{a_\text{s}^{m-1}}{m!}\left(\frac{l-s}{V_\text{g}}\right)^m.
\]
Recalling that $A_m=\frac{1}{m!}\prod_{m'=1}^m k_\text{on}^{(m')}R^{(m')}$ (Eq.~\eqref{eq:Am}), the cooperative model is equivalent to an infinite superposition of \emph{non-cooperative} multistep expressions, $\sum_{m=1}^\infty\phi_+^{(m)}$, with all recruitment rates beyond the first stage set equal: $k_\text{on}^{(m)}R^{(m)}=k_\text{self}^{(1)}R^{(1)}=a_\text{s}$ for $m>1$.

\subsection{Branchpoint drift in the cooperative model}
\label{app:cooperative-drift}
We may now compute the drift velocity of expected branching events following the same line of derivation as in Section~\ref{sec:drift}. We consider a single mother MT nucleated at $x_0$. At age $\tau$, its minus-end position and length are $X(\tau)=x_0-V_\parallel \tau$ and $L(\tau)=V_\text{g}\tau$. By Eq.~\eqref{eq phi1 self def}, the distribution of the branching complexes along the length of this MT is $\propto e^{a_\text{s}\frac{L-s}{V_\text{g}}}-1$, with center of mass given by
\[
X^{(1)} -X= \frac{\int_0^L s\left(e^{a_\text{s}\frac{L-s}{V_\text{g}}}-1\right)\, ds}{\int_0^L \left(e^{a_\text{s}\frac{L-s}{V_\text{g}}}-1\right)\, ds}. 
\]
The denominator gives
\[
\int_0^L \left(e^{a_\text{s}(L-s)/V_\text{g}}-1\right)\, ds=\frac{e^{a_\text{s} L/V_\text{g}}-1}{a_\text{s}/V_\text{g}}-L=\frac{V_\text{g}}{a_\text{s}}\sum_{m=2}^\infty\frac{(a_\text{s}L/V_\text{g})^m}{m!}.
\]
The numerator gives
\[
\int_0^L s\left(e^{a_\text{s}(L-s)/V_\text{g}}-1\right)\, ds
=
\frac{
e^{a_\text{s} L/V_\text{g}}-1-a_\text{s} L/V_\text{g}
}{
(a_\text{s}/V_\text{g})^2
}
-\frac{L^2}{2}
=\left(\frac{V_\text{g}}{a_\text{s}}\right)^2\sum_{m=3}^\infty\frac{(a_\text{s}L/V_\text{g})^m}{m!}.
\]
Hence,
\[
X^{(1)} -X = \frac{L}{3}-\frac{a_\text{s}L^2}{36V_\text{g}}+O(L^3).
\]
Substituting $X(\tau)$, $L(\tau)$ and $a_\text{s}$ yields:
\[
X^{(1)}(\tau)\approx x_0+(-V_\parallel+V_\text{g}/3)\tau-\frac{a_\text{s}V_\text{g}}{36}\tau^2.
\]
This equation reflects that when the parent MT is young and short, the drift of successive branching events follows our previous (non-cooperative) $M=1$ model. This is expected because the profile of $\phi_+^{(1)}(s)$ along the parent MT is linear in this limit. Additionally, as the parent MT ages and grows, the cooperative binding (last term on RHS) drives a \emph{leftward} drift. This is because the minus-end bias of $\phi_+^{(1)}(s)$ along the parent MT becomes increasingly nonlinear as $L$ increases.

A backward mean displacement corresponds to \(\langle X^{(1)}\rangle<x_0\), with \(\langle\cdot\rangle\) denoting an average over the parent-age distribution of branching events. As in Appendix~\ref{app:gradient-drift}, the nonlinear dependence on parent age means that evaluating \(X^{(1)}(\tau)\) at the mean age does not give the mean position. Moreover, the short-age expansion above need not hold over the ages contributing appreciably to branching. We therefore average the exact expression for \(X^{(1)}(\tau)\) over the cooperative model's parent-age distribution \(h(\tau)\), assuming spatially uniform nucleator availability. This distribution is proportional to the integrated branching complex density along the parent, multiplied by its survival probability:
\[
h(\tau)\propto e^{-\lambda\tau}\int_0^L\left(e^{a_\text{s}(L-s)/V_\text{g}} -1\right)ds \Big|_{L=V_\text{g}\tau}\propto e^{-\lambda\tau}\left(e^{a_\text{s} \tau} -1  - a_\text{s}\tau\right).
\]
The mean parent age is then
\[
\langle\tau\rangle_h=\frac{\int_0^\infty \tau h(\tau)\,d\tau}{\int_0^\infty  h(\tau)\,d\tau}=\frac{3\lambda-2 a_\text{s}}{\lambda(\lambda-a_\text{s})}.
\]
As $a_\text{s}\to 0$, the mean parent age approaches $3/\lambda$, recovering the non-cooperative $M=1$ result. Increasing cooperative recruitment shifts branching toward older parents; the mean age diverges as $a_\text{s}$ approaches $\lambda$ from below.

\begin{figure}[tp]
\begin{center}
\includegraphics[scale=1]{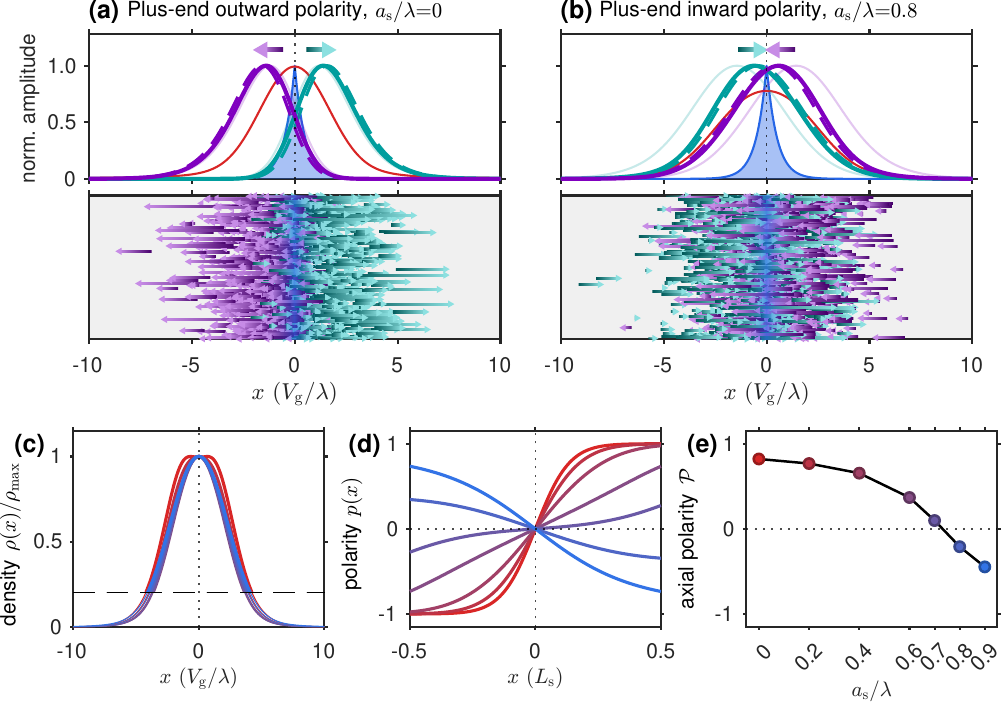}
\end{center}
\caption{{\bf The cooperative model achieves correct spindle polarity with a single intermediate species.}
All panels show steady-state solutions (snapshots at $t=200\lambda^{-1}$) with $M=1$ and $V_\parallel/V_\text{g}=0.2$ fixed, while the cooperative self-recruitment rate $a_\text{s}$ (Appendix~\ref{app:cooperative}) is varied.
\textbf{(a)} $a_\text{s}/\lambda=0$ (no self-recruitment): the system reaches plus-end outward (wrong) polarity, as in the non-cooperative $M=1$ case.
\textbf{(b)} $a_\text{s}/\lambda=0.8$ (strong self-recruitment): the system reaches plus-end inward (correct) polarity.
Top row: Ran source $s(x)$ (blue, filled) and steady-state profiles of the Ran-activated species $c(x)$ (red), MT lattice densities $\phi^{(0)}_\pm$ (solid teal/purple), and plus-end densities $\eta_\pm$ (dashed). Faint teal/purple curves show the intermediate complex density $\phi^{(1)}_\pm$.
Arrows at the top of the graphs indicate the polarity of the respective MT population; their arrowheads (plus ends) align with the peak of the plus-end densities $\eta_\pm$, and their lengths equal $\ell = V_\text{g}/\lambda$ for reference.
Bottom row: individual filament realizations (sampled from the steady-state $\psi_\pm(x,l)$; see Appendix~\ref{app:characteristics}), colored by MT polarity (teal: right-pointing; purple: left-pointing), with the Ran source density overlaid as a semi-transparent blue haze.
\textbf{(c)} Normalized MT density $\rho(x)/\rho_\text{max}$, with $\rho = \phi^{(0)}_+ + \phi^{(0)}_-$, for $a_\text{s}/\lambda$ from $0$ to $0.9$ (red to blue); thick portion marks the spindle interior ($\rho > 0.2\,\rho_\text{max}$).
\textbf{(d)} Local polarity $p(x) = (\phi^{(0)}_+-\phi^{(0)}_-)/\rho$ within the spindle interior, $x$ normalized by spindle length $L_\text{s}$.
\textbf{(e)} Axial polarity $\mathcal{P}$ vs.\ $a_\text{s}/\lambda$; $\mathcal{P}>0$ indicates wrong polarity (plus ends outward), $\mathcal{P}<0$ correct polarity (plus ends inward). The reversal occurs slightly past $a_\text{s}=0.7\,\lambda$, roughly consistent with the analytical drift-reversal condition $a_\text{s}>2/3\,\lambda$ derived under the uniform nucleator availability assumption in Appendix~\ref{app:cooperative-drift}.
}
\label{fig:cooperative-fields}
\end{figure}

The mean branching position at age $\tau$ is given by $X^{(1)}(\tau)=x_0-V_\parallel \tau +\bar{s}(\tau)$, where $\bar{s}(\tau)$ denotes the mean branch distance from the parent's current minus end at age $\tau$,
\[
\bar s(\tau)=\frac{
\displaystyle\int_0^{V_\text{g}\tau}
s\left(e^{a_\text{s}(V_\text{g}\tau-s)/V_\text{g}}-1\right)\,ds
}{
\displaystyle\int_0^{V_\text{g}\tau}
\left(e^{a_\text{s}(V_\text{g}\tau-s)/V_\text{g}}-1\right)\,ds
}=   \frac{V_\text{g}}{a_\text{s}}
\frac{e^{a_\text{s}\tau}-1-a_\text{s}\tau-\tfrac12(a_\text{s}\tau)^2}
{e^{a_\text{s}\tau}-1-a_\text{s}\tau}.
\]
Note how $h(\tau)$ is proportional to the denominator in $\bar s(\tau)$ times $e^{-\lambda\tau}$. The mean $\bar{s}$ with respect to $h(\tau)$ is then
\[
\langle \bar{s}(\tau)\rangle_h =\frac{\displaystyle\int_0^\infty \bar s(\tau)h(\tau)\,d\tau}{\displaystyle\int_0^\infty h(\tau)\,d\tau} = \frac{V_\text{g}}{a_\text{s}} \frac{\displaystyle\int_0^\infty e^{-\lambda\tau}\left(e^{a_\text{s}\tau}-1-a_\text{s}\tau-\tfrac12(a_\text{s}\tau)^2\right)\,d\tau
}{\displaystyle\int_0^\infty e^{-\lambda\tau}\left(e^{a_\text{s}\tau}-1-a_\text{s}\tau\right)\,d\tau}= \frac{V_\text{g}}{a_\text{s}}
\frac{a_\text{s}^3/[\lambda^3(\lambda-a_\text{s})]}
{a_\text{s}^2/[\lambda^2(\lambda-a_\text{s})]}
= \frac{V_\text{g}}{\lambda}.
\]
We therefore find that the mean branch distance from the parent's minus end is independent of the cooperative recruitment rate $a_\text{s}$, and is exactly equal to the typical MT length $V_\text{g}/\lambda$, as in the non-cooperative case.

In summary, the mean position of branching events in the cooperative model is
\begin{equation}
\langle X^{(1)}\rangle_h = x_0-V_\parallel\frac{3\lambda-2a_\text{s}}
{\lambda(\lambda-a_\text{s})}+\frac{V_\text{g}}{\lambda}.
\label{eq:mean-Xn-cooperative}
\end{equation}
This result is exact under the uniform-nucleator assumption for
$0<a_\text{s}<\lambda$, with the non-cooperative case recovered
as $a_\text{s}\to0$. The criterion $\langle X^{(1)}\rangle_h < x_0$ for backward drift is thus satisfied when
\begin{equation}
\frac{V_\parallel}{V_\text{g}} > \frac{\lambda-a_\text{s}}{3\lambda-2a_\text{s}}.
\label{eq:cooperative-drift-condition}
\end{equation}
Alternatively, for $V_\parallel/V_\text{g}<1/2$ and with $0<a_\text{s}<\lambda$ as the control parameter, backward drift is achieved in the regime
\begin{equation}
\frac{V_\text{g}-3V_\parallel}
{V_\text{g}-2V_\parallel} \lambda<a_\text{s}<\lambda.
\label{eq:as-regime}
\end{equation}
Substituting the reference ratio $V_\parallel/V_\text{g}=1/5$ gives
\begin{equation}
\tfrac{2}{3}\lambda < a_\text{s} < \lambda.
\label{eq:as-regime-ref}
\end{equation}
The cooperative mechanism can therefore produce leftward drift with measured velocities, but only when the self-recruitment rate $a_\text{s}$ is of order the stability bound $\lambda$. Our simulations of this model likewise yield inward polarity ($\mathcal{P}<0$) at sufficiently strong cooperative recruitment; see Fig.~\ref{fig:cooperative-fields}. We note that the numerical transition in axial polarity need not coincide precisely with the analytical threshold for backward mean displacement, which assumes spatially uniform nucleator availability.  

The two assembly mechanisms analyzed in this paper (multistep and cooperative) may also act in concert: cooperative self-recruitment at even one or two assembly stages could supply part of the requisite minus-end bias, reducing the number of sequential rate-limiting steps needed to achieve correct polarity. At the measured velocity ratio, backward drift requires branching to be sufficiently biased toward the parent MT's minus end. Both mechanisms produce this bias through a strong increase in branching-complex density with the age of the MT lattice.

\section{Supplementary movies}
\label{app:movies}
Movies of the stochastic filament simulations (detailed in Appendix~\ref{app:gillespie}) illustrate the results from the continuum model. Each movie has four stacked panels from top to bottom showing the time evolution of: the continuum nucleation rates $n_\pm(x,t)$; the minus ends of the stochastic filaments; their plus ends; and the filaments themselves. Right-pointing MTs are shown in cyan and left-pointing MTs in magenta. Frames are spaced by $\Delta t_\text{save}=0.1\,\lambda^{-1}$ and span $0\le t\le 200\,\lambda^{-1}$. The movies are available from the corresponding author upon request.

\bigskip
\refstepcounter{movie}\label{mov:M0-eps02}\noindent\textbf{Movie~\themovie.} Single-step model at the measured velocity ratio ($M=0$, $V_\parallel/V_\text{g}=0.2$), which relaxes to an inverted, plus-end outward polarity (see Fig.~\ref{fig:dynamics} and Fig.~\ref{fig:M0-fields}a).

\bigskip
\refstepcounter{movie}\label{mov:M0-eps06}\noindent\textbf{Movie~\themovie.} Single-step model at a higher velocity ratio ($M=0$, $V_\parallel/V_\text{g}=0.6$), which relaxes to the correct, plus-end inward polarity (see Fig.~\ref{fig:M0-fields}b).

\bigskip
\refstepcounter{movie}\label{mov:M5-eps02}\noindent\textbf{Movie~\themovie.} Multistep model at the measured velocity ratio ($M=5$, $V_\parallel/V_\text{g}=0.2$), which relaxes to the correct, plus-end inward polarity (see Fig.~\ref{fig:dynamics-M5} and Fig.~\ref{fig:Mstep-fields}b).

\bigskip
\refstepcounter{movie}\label{mov:M1-eps02}\noindent\textbf{Movie~\themovie.} Two-step model at the measured velocity ratio ($M=1$, $V_\parallel/V_\text{g}=0.2$), which relaxes to an inverted, plus-end outward polarity (see Fig.~\ref{fig:Mstep-fields}a).

\end{appendices}

\bibliographystyle{unsrt}
\bibliography{refs}

\clearpage

\end{document}